\documentclass{aa}  
\usepackage{graphicx}
\usepackage{txfonts}
\usepackage{tabularx}
\usepackage{amsfonts}
\usepackage{bbold}
\usepackage{color}
\usepackage{transparent}
\usepackage{hyperref}
\usepackage{transparent}
\usepackage{rotating}
\usepackage{caption}
\usepackage{subcaption}

\usepackage{graphicx}	
\usepackage{amsmath}	
\usepackage{amssymb}	
\usepackage{tablefootnote}
\usepackage[flushleft]{threeparttable}
\usepackage{dblfloatfix} 
\usepackage{siunitx}
\usepackage{array}

\usepackage{physics}

\usepackage{multirow}
\usepackage{diagbox}

\usepackage{natbib}
\bibpunct{(}{)}{;}{a}{}{,} 

\newcommand{\sgx}{SgXB\xspace}
\newcommand{\sgxs}{SgXBs\xspace}

\newcommand{\sfxt}{SFXT\xspace}
\newcommand{\sfxts}{SFXTs\xspace}

\newcommand*{\hmxb}{HMXB\@\xspace}
\newcommand*{\hmxbs}{HMXBs\@\xspace}

\newcommand*{\ns}{NS\@\xspace}

\newcommand*{\bh}{BH\@\xspace}

\newcommand*{\eg}{e.g.,\@\xspace}
\newcommand*{\ie}{i.e.,\@\xspace}

\newcommand{\mystar}{{\fontfamily{lmr}\selectfont$\star$}}
\newcommand*{\msun}{$M_{\odot}$\@\xspace}
\newcommand*{\los}{LOS\xspace}

\newcommand*{\blue}{\textcolor[rgb]{0,0,0}}
\newcommand*{\red}{\textcolor[rgb]{0,0,0}}
\newcommand*{\orange}{\textcolor[rgb]{0,0,0}}

\usepackage[dvipsnames]{xcolor}

\makeatletter
\newcommand{\thickhline}{%
    \noalign {\ifnum 0=`}\fi \hrule height 1pt
    \futurelet \reserved@a \@xhline
}
\makeatother

\begin{document}

   \title{Radiography in high mass X-ray binaries}

   \subtitle{Micro-structure of the stellar wind through variability of the column density}

   \author{I. El Mellah
          \inst{1}
          \and
          V. Grinberg
          \inst{2}
         \and
          J.~O.~Sundqvist
          \inst{3}          
          \and
          F.~A.~Driessen
          \inst{3}   
          \and
          M.~A.~Leutenegger
          \inst{4}
          }

   \institute{Centre for mathematical Plasma Astrophysics, 
   			 Department of Mathematics, KU Leuven, 
   			 Celestijnenlaan 200B, B-3001 Leuven, Belgium\\
              \email{ileyk.elmellah@kuleuven.be}
         \and
         	Institut f\"ur Astronomie und Astrophysik, Universität T\"ubingen, Sand 1, 72076 T\"ubingen, Germany 
         \and
         	Institute of Astronomy, KU Leuven,
              Celestijnenlaan 200D, B-3001 Leuven, Belgium
        \and
            NASA Goddard Space Flight Center, 8800 Greenbelt Rd., Greenbelt, MD 20771, USA
            \label{inst:gsfc}
    }
   \date{Received ...; accepted ...}

 
  \abstract
   {In high mass X-ray binaries, an accreting compact object orbits a high mass star, which loses mass through a dense and inhomogeneous wind.}
   {Using the compact object as an X-ray backlight, the time variability of the absorbing column density in the wind can be exploited in order to shed light on the micro-structure of the wind and obtain unbiased stellar mass loss rates for high mass stars.}
   {We developed a simplified representation of the stellar wind where all the matter is gathered in spherical "clumps" that are radially advected away from the star. This model enables us to explore the connections between the stochastic properties of the wind and the variability of the column density for a comprehensive set of parameters related to the orbit and to the wind micro-structure, such as the size of the clumps and their individual mass. In particular, we focus on the evolution with the orbital phase of the standard deviation of the column density and of the characteristic duration of enhanced absorption episodes. Using the porosity length, we derive analytical predictions and compare them to the standard deviations and coherence time scales that were obtained.}
   {We identified the favorable systems and orbital phases to determine the wind micro-structure. The coherence time scale of the column density is shown to be the self-crossing time of a single clump in front of the compact object. We thus provide a procedure to get accurate measurements of the size and of the mass of the clumps, purely based on the observable time variability of the column density.}
   {The coherence time scale grants direct access to the size of the clumps, while their mass can be deduced separately from the amplitude of the variability. We further show how monitoring the variability at superior conjunctions can probe the onset of the clump-forming region above the stellar photosphere. If the high column density variations in some high mass X-ray binaries are due to unaccreted clumps which are passing by the line-of-sight, this would require high mass clumps to reproduce the observed peak-to-peak amplitude and coherence time scales. These clump properties are marginally compatible with the ones derived from radiative-hydrodynamics simulations. Alternatively, the following components could contribute to the variability of the column density: larger orbital scale structures produced by a mechanism that has yet to be identified or a dense environment in the immediate vicinity of the accretor, such as an accretion disk, an outflow, or a spherical shell surrounding the magnetosphere of the accreting neutron star.}
   


   \maketitle
%

\section{Introduction}

Massive stars live a forceful life during which they shape galaxies with their mechanical, radiative, and chemical feedback. Evolutionary pathways of massive stars are largely controlled by their line-driven winds, with \red{typical} mass loss rates of 10$^{-6}\text{\msun}\cdot$yr$^{-1}$ and terminal wind speeds on the order of 1\,000 km$/$s \citep{Puls2008}. The flow is provided with net outward momentum thanks to the resonant line absorption of ultraviolet (UV) stellar photons by partly ionized metal ions present in the outer envelope of the star. As the wind accelerates, it keeps tapping higher energy photons thanks to the Doppler shift \citep{Lucy1970,Castor1975}. 

\red{Originally, these line-driven winds were treated as smooth, steady-state outflows without any internal shocks or overdensity. However, it has been know for a long time now that they are, in fact, highly structured and variable on a broad range of spatial and temporal scales (for overviews, see \citealt{Hamann08}; \citealt{Sundqvist12c}; \citealt{Puls15}).}  
Indeed, the dependence of the acceleration on the velocity profile itself causes the emergence of a self-growing feedback loop, the line-deshadowing instability, which is responsible for internal shocks and the formation of overdense regions or "clumps" \citep{Owocki1984,Owocki1988,Feldmeier1995}.
\red{This clumping can then severely affect the radiative transfer} computations used to \red{derive synthetic observables; for example, neglecting wind clumping leads to empirically inferred mass-loss rates that may differ by an order of magnitude for the same star, depending on which observational diagnostic is used to estimate the mass loss \citep{Fullerton06}. Typically, mass-loss rates derived from diagnostics with opacities depending on the square of the density (\eg the H$\alpha$ line and thermal radio emission) are overestimated if clumping is neglected, whereas those derived from diagnostics depending linearly on density (usually UV resonance lines) can be underestimated if the light-leakage effects of a wind that is porous in velocity-space are not properly taken into account \citep[\eg][]{Sundqvist18}.} X-ray emission line shapes from single stars offer an alternative mass-loss rate diagnostic based on continuum photoelectric opacity which is linear in density and insensitive to velocity field inhomogeneity \citep{Owocki2001, Cohen2014a}. This diagnostic might still be compromised by geometrical porosity \citep{Feldmeier2003, Sundqvist2012}, although observational constraints disfavor the extreme degree of wind structure that would be needed for this \citep{Leutenegger2013}. Due to the faintness of X-ray emission from single stars and the modest size of existing X-ray telescopes, this diagnostic is currently limited to a handful of nearby objects.

High-mass X-ray binaries (\hmxbs) can shed a unique light upon the \red{mass loss and} micro-structure of line-driven winds. In Supergiant X-ray binaries (\sgxs), a subclass of \hmxbs, a compact object, generally a neutron star (\ns), sometimes a black hole (\bh), orbits a blue supergiant and accretes part of its winds \citep{Walter15,Martinez-Nunez2017}. The material captured heats up as it falls into the gravitational potential well and emits copious amounts of X-rays from the immediate vicinity of the compact object. Two methods have been proposed in order to use the X-ray point source as a probe and bring constrains on the wind clumpiness independent from the ones obtained with diagnostics for isolated massive star \citep{Martinez-Nunez2017}. 

In the first method, if the episodic enhancements of the intrinsic X-ray luminosity observed in \sgxs are due to the direct capture of a clump, the amount of mass in a clump and the number of clumps per unit volume could be estimated \citep{IntZand2005}. This idea was applied both in persistent \sgxs \citep[also called classic \sgxs, with low intrinsic variability,]{Ducci2009} and in supergiant fast X-ray transients \citep[\sfxts,][]{Bozzo_2017a}. In a 1D framework, \citet{Oskinova2012} have calculated synthetic X-ray lightcurves that would result from this approach and compared the expected variability to the observed. This method presupposes that the variability induced by the wind micro-structure remains unaltered from the orbital scale all the way down to the X-ray emitting region (either the \ns surface or the inner rim of the accretion disk surrounding the \bh). Under this assumption, \cite{Oskinova2012} obtained a peak-to-peak amplitude of the X-ray luminosity several orders of magnitude higher than what observed in \sgxs. However, additional structures form, such as a hydrodynamics bow shock around the accretor, which have been shown to lower the variability compared to a purely ballistic model of the clump capture \citep{ElMellah}. The amount of material involved in an observed flare is thus an upper limit on the mass of a clump: a flare might be triggered by the capture of a clump, as noticed by \cite{Bozzo2015}, but likely involves additional material which has been previously piling up at the outer rim of the \ns magnetosphere due to the magneto-centrifugal gating mechanism \citep[or propeller effect,][]{Illarionov1975,Grebenev2007,Bozzo2008}, in a quasi-spherical hot shell on top of the \ns magnetosphere \citep{Shakura2013b} or in a disk-like structure \citep{ElMellah2018,Karino2019}. Also, \cite{Hainich2020} showed that there was no systematic difference between the stellar properties in classic \sgxs and in \sfxts, hinting at similar wind micro-structures \citep{Driessen2019}. Finally, \cite{Pradhan2017} concluded that the higher variability of the emission in \sfxts compared to \sgxs was likely due to mechanisms inhibiting accretion near the accretor, rather than direct clump capture, although flares in \sfxts have recently been found by \cite{Ferrigno2020} to be associated with massive structures approaching the accretor. To summarize, X-ray variability is, at best, an indirect tracer of the clumpiness of the wind because of intermediate regions where the clumps are mixed and where other instabilities can kick in, in particular those associated to the different accretion regimes.

In a second method, one can make use of the highly variable absorption local to the HMXBs.
If the variations in the absorbing column density are due to unaccreted clumps passing through the line-of-sight (\los), the accretor could play the role of a steady backlight whose shimmering betrays the stochastic properties of the wind it is embedded into \citep{Oskinova2012,Grinberg2015}. Time resolved spectroscopy enables us to monitor the absorbing column density directly \citep[\eg][]{Haberl1989,Martinez-Nunez2014,Grinberg2015}. Shorter time scales may be indirectly accessible via changes in X-ray color \citep[\eg][]{Grinberg2017,Bozzo_2017a}, especially when using multiple energy bands and working with color-color diagrams \citep{Nowak_2011a,Hirsch2019}.

\cite{Grinberg2015} laid the foundations of a model of the clumps to relate the observed dips in the X-ray light curve of the \hmxb Cygnus X-1 to the micro-structure of the wind. They assumed spherical clumps emitted by the donor star on radial trajectories. However, their exploration was driven by the measurements available using the particular set of observations discussed in their work and limited to the well known parameters of one particular system whose observable properties their model aimed to reproduce. In parallel, new insights have recently been provided on the clump properties thanks to multidimensional radiative hydrodynamics simulations \citep{Sundqvist2017} and on the velocity profiles thanks to 1D radiative transfer calculations in a steadily expanding atmosphere in nonlocal thermodynamical equilibrium \citep{Sander2017b}. These results drove us into expanding on the first ansatz introduced by \cite{Grinberg2015}. We aim at designing a minimal representation of a clumpy wind suitable to grasp the complexity of the flow while still including analytically derivable predictions to prove its robustness. We want to determine whether, in realistic conditions, we can use the median value, the amplitude variations and the coherence time scales of the column density as a function of the orbital phase in order to draw conclusions on the wind micro-structure, in spite of the large number of clumps and of the diversity of clump properties along the \los. To do so, we carry out a parametric study based on physically-motivated ranges for parameters controlling the orbit, the wind velocity profile and the mass and radius of the clumps. We discuss the observational consequences to be expected according to this theoretical framework but we postpone to later a detailed conversation about the instrumental capabilities and limitations of the current and upcoming X-ray satellites. Also, we focus on the impact of the clumps and discard other sources of column density variability such as larger scale structures produced by the orbital motion for instance.

In section\,\ref{sec:model}, we describe the ingredients of the model we developed and the numerical procedure we follow to compute the column density produced by clumps in number high enough to match the expected number of clumps around massive stars in \hmxbs. In section\,\ref{sec:analytics}, we present the analytic derivation of the column density variance that we confront to the numerical results in section\,\ref{sec:results}, where we also report on the characteristic time scales of absorption episodes and the dependences with respect to the parameters of the model. In section\,\ref{sec:diagnostics}, we present how these results can be used to constrain the wind micro-structure from observations, while the limitations of our framework are acknowledged in section\,\ref{sec:limitations}. Finally, we summarize our findings in section\,\ref{sec:summary}.

\section{Model}
\label{sec:model}


\subsection{Wind velocity profile}
\label{sec:iso_wind}


Hot stars provide their outer envelopes with plenty of high energy photons able to drive a wind. Most of the predictions from the seminal papers on the theory of radiation-driven winds by \cite{Lucy1970} and \cite{Castor1975} have been qualitatively confirmed by observations of isolated hot stars, on the main sequence but also evolved \citep{Puls2006}. In the Sobolev approximation, the radial optical depth can be written as a local quantity modulated by the radial velocity gradient \citep{Sobolev1960}. The wind velocity profile can then be determined analytically and it has been common since then to prescribe, for the dynamics of the wind, a radial velocity profile which obeys the $\beta$-law, a generalization of the first analytic formula derived \citep{Puls2008}. Its two parameters are the terminal wind speed $\varv_{\infty}$, highly supersonic, and a $\beta$ exponent which quantifies how quickly the terminal speed is reached :
\begin{equation}
\label{eq:beta_law}
\varv(r)=\varv_{\infty}\left(1-R_{\text{\mystar}}/r\right)^{\beta},
\end{equation}
where $r$ is the distance to the center of the star and $R_{\text{\mystar}}$ is the stellar radius. The larger $\beta$, the more gradual the acceleration. 

In Vela X-1, observations of UV spectral lines led \cite{Gimenez-Garcia2016} to propose, for the wind of the donor star HD 77581 of spectral type B0.5Iae \citep{VanKerkwijk1995}, a $\beta$-law with $\varv_{\infty}\sim 700$ km$/$s and $\beta=1$. 
\red{Later on, \cite{Sander2017} computed a hydrodynamic steady-state structure for the wind stratification of the donor star in Vela X-1 based on a nonlocal thermodynamical equilibrium computation of the population on involved energy levels.} Between the photosphere and the compact object and for our current purpose, the velocity profile \red{predicted} by \cite{Sander2017} can be reasonably well fit by a $\beta$-law with $\varv_{\infty}\sim 500$ km$/$s and $\beta\sim 2.3$. In our simulations, we consider two different values for $\beta$: 0.5, representative of a quickly accelerating wind, and 2, for a slowly accelerating wind. The terminal wind speed in Vela X-1 lies on the lower end of what is usually found in \hmxbs (see Table\,\ref{tab:vinf} for a few other systems), which leads to a wind focused toward the accretor and susceptible to form transient wind-captured disks \citep{ElMellah2018}, a prediction recently corroborated by observations \citep{Liao2020}. Given how high the terminal wind speed usually is compared to the orbital speed in \sgxs, where the orbital period is typically of at least a few days, the wind ballistic streamlines are not significantly curved \citep{ElMellah2016a}. In several \hmxbs, the gravitational influence of the compact object has little impact at the orbital scale and the wind \red{should be} essentially radial and isotropic around the donor star.

\setlength{\tabcolsep}{3pt}
\begin{table}[]
\caption{Estimated terminal wind speeds from observations in a few \sgxs.}
\label{tab:vinf}
\begin{tabular}{lll}
\hline\hline
\sgxs & $v_{\infty}$ & Reference \\ 
\hline 
Vela X-1 & 700 km$/$s & \cite{Gimenez-Garcia2016} \\ 
Cygnus X-1 & 2\,100 km$/$s & \cite{Herrero1995} \\ 
4U 1700$-$377 & 1\,700 km$/$s & \cite{VanLoon2001} \\ 
IGR J00370$+$6122 & 1\,100 km$/$s & \cite{Hainich2020} \\
\hline
\end{tabular}
\end{table}
\setlength{\tabcolsep}{6pt}

\subsection{Clumps}
\label{sec:clumps}
In the two-zones model we work with, all the \red{wind mass} is contained in spherical clumps, each being uniformly dense, with 
\red{an effectively} void inter-clump environment. \blue{This approximation is justified given the observational findings that most of the mass of the wind material in the two prototypical HMXBs Cyg X-1 and Vela X-1 is concentrated in a small fraction of the volume \citep{Sako1999,Rahoui:2011bo}}. The clumps are carried with the wind, on purely radial orbits. They do not merge with each other, which enables us to use the continuity equation to deduce $n$, the number of clumps per unit volume at a certain distance $r$ from the stellar center (not to be confused with the particle number density of the smooth wind):
\begin{equation}
n=\frac{\dot{\mathcal{N}}}{4\pi r^2 v},
\label{eq:mdot_ndot}
\end{equation}
where $\dot{\mathcal{N}}$ is the rate at which clumps are emitted from the stellar photosphere, given by $\dot{M}/m$. $\dot{M}$ is the stellar mass loss rate and $m$ is the mass of one clump, assumed independent of the distance to the star. All the clumps have the same mass $m$ and the same radius $R_{cl,2}$ at two stellar radii, which are set as input parameters of the model. The evolution of the clump radius with the distance to the star can be set based on observational or physical insights concerning how clumpy the wind is. The latter is quantified through the clumping factor $f$ or the volume filling factor $f_{\rm vol}$:
\begin{equation}
\begin{cases}
\red{f = \langle \rho^2 \rangle / \langle \rho \rangle ^2} \\
f_{\rm vol} = \rho / \rho_{cl},
\end{cases}
\label{eq:factors}
\end{equation}
where $\rho$ is the smooth wind density and $\rho_{cl}$ is the mass density of a clump. \red{When the inter-clump environment is empty, $\langle \rho \rangle = \rho_{\rm cl} f_{\rm vol}$ but $\langle \rho^2 \rangle = \rho_{\rm cl}^2 f_{\rm vol}$, such that $\langle \rho^2 \rangle / \langle \rho \rangle^2 = 1/f_{\rm vol} = f$.} 



In the case of a uniform clumping factor through the wind, the evolution of the clump radius $R_{cl}$ with the distance $r$ to the stellar center is best represented by:
\begin{equation}
\label{eq:lorenzo_ext}
    R_{cl} \propto (vr^2)^{1/3}.
\end{equation}
It is also the profile which ensures that the mass density of constant mass clumps evolves as the density of a smooth wind \citep{Ducci2009}. Alternatively, \red{it has been proposed based on the observed and simulated radial stratification of $f$ in O-stars
\citep{Sundqvist13} that in a simple model such as here the clumps might expand according to \citep[see the Appendix A in ][]{Sundqvist2012}:}
\begin{equation}
\label{eq:jon_ext}
    R_{cl} \propto r.
\end{equation}
Although the exact clump radius expansion law is unknown, radiative hydrodynamics simulations indicate that the radius of the clumps tend to increase as they flow away from the star \citep{Sundqvist2017}. Hereafter, these two expansion laws are referred to as smoothly expanding clumps and as linearly expanding clumps respectively.

\red{The absorption produced by clumps can be quantified through the porosity length, the effective clump column density divided by the mean density \citep{Owocki2006}.} The porosity length $h$ can be interpreted as the distance in a smooth wind along which the column density is the same as the column density of one clump averaged over all possible impact parameters $p$:
\begin{equation}
\label{eq:porosity_length}
    \red{\rho _{cl} \left(\frac{1}{2R_{cl}}\int_{-R_{cl}}^{R_{cl}}2\sqrt{R_{cl}^2-p^2}\,dp \right)=\rho _{cl}\frac{\pi}{2}R_{cl} = \langle \rho \rangle h} ,
\end{equation}
which means that the porosity length profile can be written as a function of the clumping factor $f$:
\begin{equation}
h=\frac{\pi}{2} f R_{cl}.    
\end{equation}

\begin{figure}
\centering
\includegraphics[width=0.99\columnwidth]{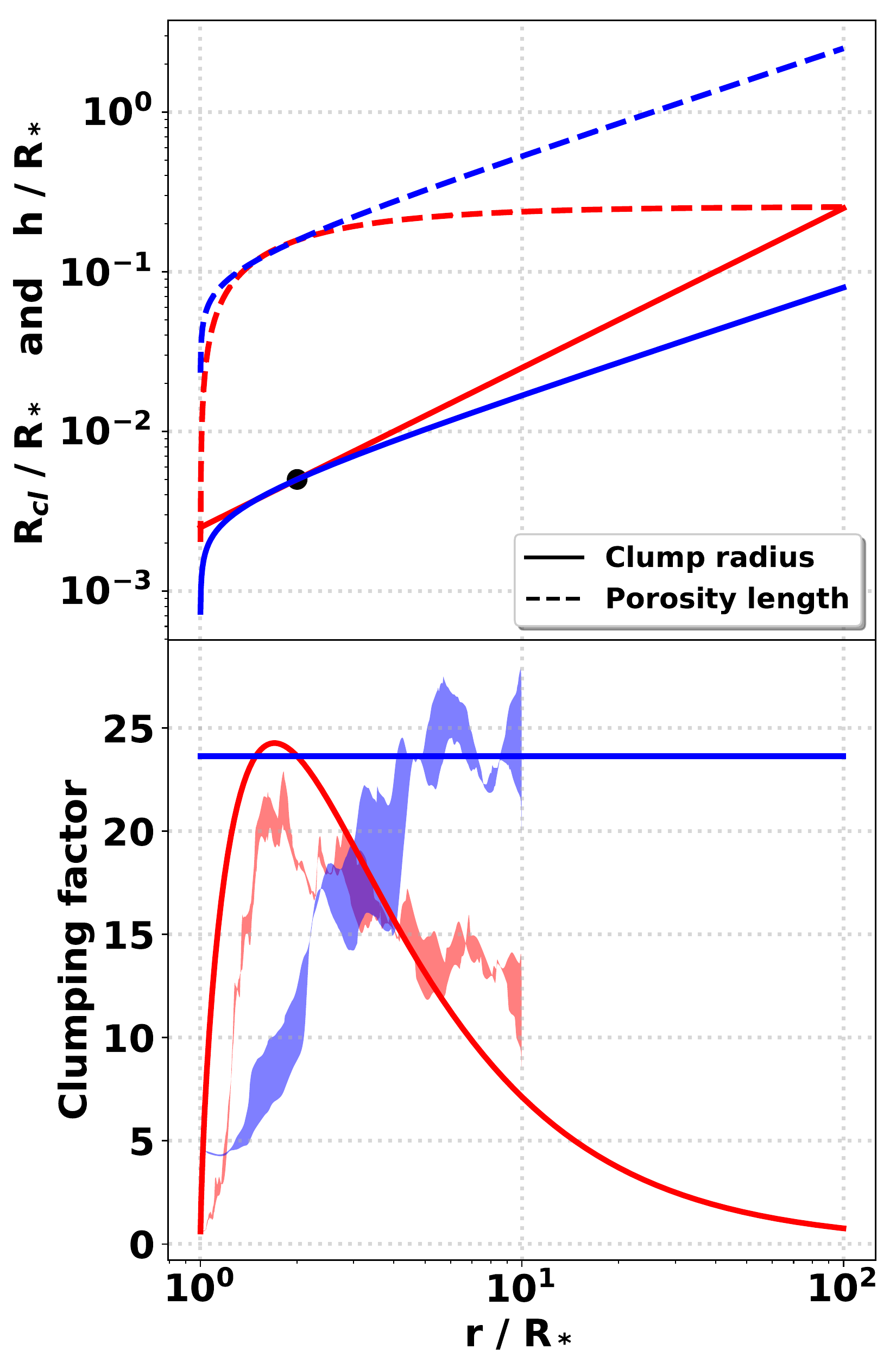}
\caption{(top panel) Clump radius (solid lines) and porosity length (dashed lines) in units of $R_{*}$ as a function of the distance to the stellar center. In blue (resp. in red) are represented the profiles for the clump radius expansion law\,\eqref{eq:lorenzo_ext} (resp.\,\eqref{eq:jon_ext}). The dimensionless clump radius at $r=2$, indicated with a black dot, was set to 0.005, and the dimensionless clump mass to 4$\cdot$10$^{-7}$ (see section\,\ref{sec:phys_par} for normalization units). (bottom panel) Clumping factor profiles for both expansion laws. The \orange{red} (resp.~\orange{blue}) shaded region represents the shape of the clumping factor profile obtained for O (resp. B) supergiants by \cite{Driessen2019}. \orange{Spectroscopically, these OB supergiants can be considered representing typical early- and middle-type stars.}
}
\label{fig:porosity}
\end{figure}

The profiles for the clump radius, the porosity length, and the clumping factors are shown in Fig.~\ref{fig:porosity} for expansion laws in Eq.~\eqref{eq:lorenzo_ext} and~\eqref{eq:jon_ext} (in blue and red, respectively). Linearly expanding clumps lead to a clumping factor peaking at $r=(\beta+1)R_*$ while in a wind with smoothly expanding clumps, the clumping factor is uniform. Additionally included in the bottom panel in Figure\,\ref{fig:porosity} are shaded regions which represent typical clumping factor profiles obtained from 1D line-driven instability simulations of two O supergiants and two B supergiants \citep{Driessen2019}. Observations of isolated \red{O stars} yield clumping factors which qualitatively agree with the ones computed in these simulations \citep{Puls2006,Najarro2011, Sundqvist13}. We point out that we are not aiming here to obtain absolute clumping factors but rather focus on the comparison with the formalism developed in this paper. Indeed, variations in clumping factor are to be expected between one-dimensional and multidimensional line-driven instability simulations \citep{Sundqvist2017}. The latter shows that clumping factors are reduced with respect to one-dimensional simulations because in 1D, the flow is unable to circumvent clumps formed in the simulation (really shells in one dimension). Accelerated interclump matter will flow into dense, clumpy material which artificially increases the clump density $\rho_{cl}$ and, hence, the clumping factor. However, in Figure\,\ref{fig:porosity}, the comparison between the profiles computed in numerical simulations (shaded regions) and those computed based on the analytic description of the clump radius expansion models (solid lines) indicates that the two simple expansion laws described above are still realistic for the wind of O and B supergiants and that the corresponding clumping factor profiles follow reasonably well the trends derived by the simulations.

Simulations and observations further suggest that clumps masses (resp. radii) lie in-between a few 10$^{17}$\,g and 10$^{19}$\,g (resp. 0.5\% and 8\% of the stellar radius), which, after considering realistic normalization in \sgxs, motivates our choice of dimensionless parameters (see Table\,\ref{tab:params} and Table\,\ref{tab:norm}).

On purpose, we compute the variability of the total column density without mentioning the X-ray luminosity itself, the optical depth or the extinction coefficient since the model we develop does not include energy-dependent opacity. In follow up work, we plan to address the question of the ionization structure of the wind, with a proper treatment of the X-ray ionizing feedback from the X-ray source. The present work concerns only the total column density, that is to say the amount of material integrated along the \los, not the absorbing column density deduced from fits of the continuum in X-ray spectra. Changes of the ionization parameter and thus, of the opacity, along the \los could induce significant variations in the absorbing column density while leaving the total column density unchanged (see also discussion in Section\,\ref{sec:ionizing_feedback}).



\subsection{Orbiting X-ray source}
\label{sec:XXX}


In \hmxbs, most of the X-rays we observe originate from close to the \ns magnetic poles or from close to the innermost regions of the accretion disk surrounding the \bh, depending on the nature of the accretor. At the orbital scale, these regions of a few tens to hundreds km are so small that the accretor can be represented as an orbiting X-ray point source in our model. 

\begin{figure}
\begin{subfigure}{.5\textwidth}
\centering
\includegraphics[width=0.99\columnwidth]{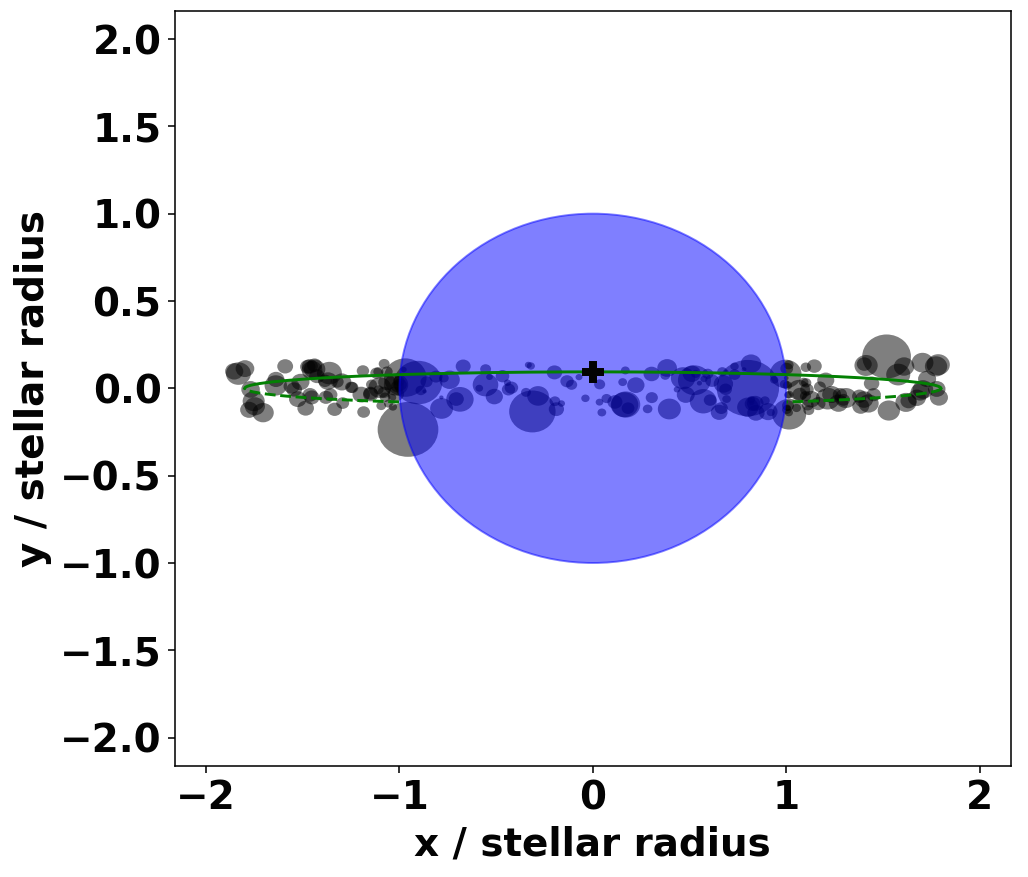}
\end{subfigure}
\phantom{p}\\
\begin{subfigure}{.5\textwidth}
\centering
\includegraphics[width=0.99\columnwidth]{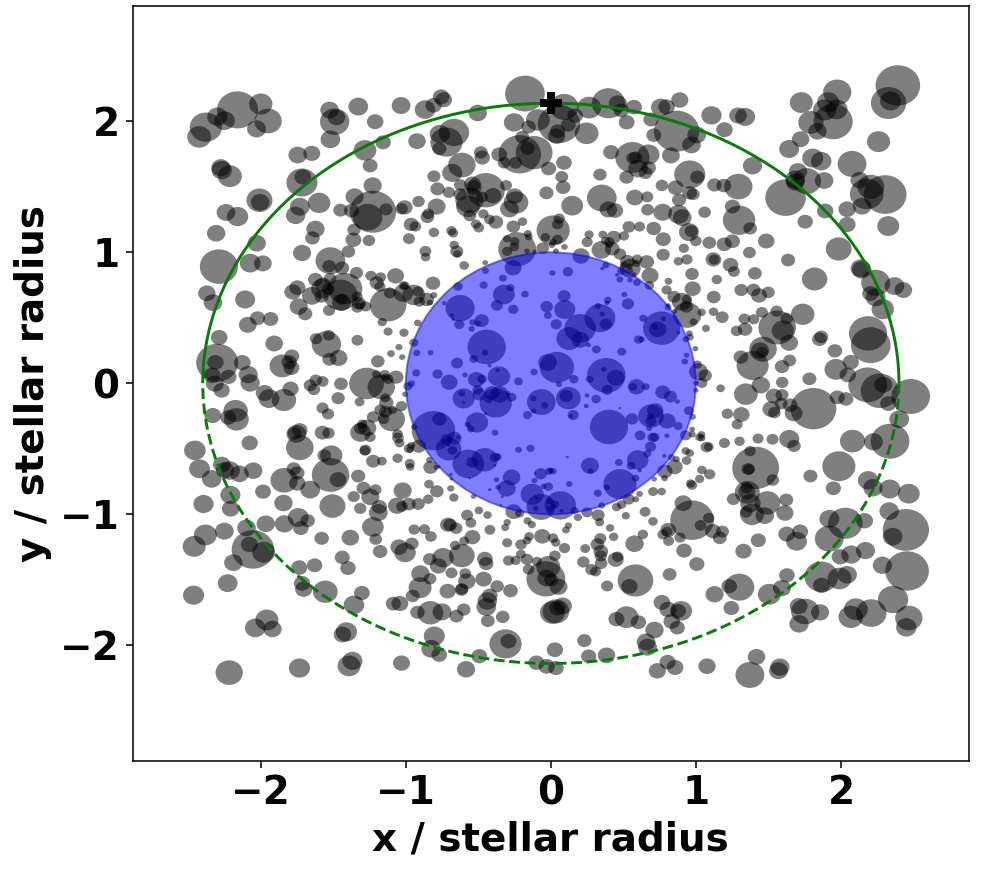}
\end{subfigure}
\caption{Representation of the orbit (in green) of the compact object around its stellar companion (in blue) for Vela X-1 (top panel) and Cygnus X-1 (bottom panel) as seen from Earth, where the eccentricity of the orbit has been neglected. Only a fraction of the clumps susceptible to contribute to the column density along the orbit have been represented (in semi-opaque grey). The black cross indicates the point of inferior conjunction on the orbit.}
\label{fig:big_picture}
\end{figure}

In classic \sgxs, the eccentricity of the orbit of the compact object is low \citep{Walter15} and so is the orbital separation, typically within $\sim$1.5--3 stellar radii, motivating our choice of orbital separations. The present model could be extended to include any type of predefined orbit but for the sake of simplicity, we only treat circular orbits (see Sect.~\ref{sec:ecc_orbits} for a discussion on eccentric orbits). The orbit is entirely determined by two parameters: the orbital separation $a$ and the inclination angle $i$ between the line of sight and the orbital angular momentum axis. In Fig.~\ref{fig:big_picture} we show two classic  \hmxbs, Vela X-1 and Cygnus X-1 as seen from Earth. The \ns-hosting Vela X-1 is essentially edge-on  as it is an eclipsing \hmxb, that is it has a high inclination angle \citep{Joss1984}. The \bh-hosting Cygnus X-1 is much more face-on, with a low inclination angle of $\sim$27$^\circ$ \citep{Orosz2011}.

\subsection{Physical parameters of the model}
\label{sec:phys_par}

Our model includes a very limited number of six degrees of freedom to evaluate the impact of the wind micro-structure on the time variability of the column density. The orbit is determined by the orbital separation $a$, the orbital inclination $i$ and the orbital period $P$. The clump properties are the mass and the radius at 2$R_*$. The dynamics of the wind is set by the $\beta$-exponent. The terminal wind speed $v_{\infty}$ is used as the normalization quantity for speeds and thus, does not affect the shape of the numerical solutions. The stellar radius $R_*$ is the normalization of the lengths and with the stellar mass loss rate $\dot{M}$ as an additional normalization unit, all our results can be scaled to physical units. The values for the six shape parameters we explore are given in Table\,\ref{tab:params} and correspond to 600 different configurations. Among these 600 configurations, a few correspond approximately to the values in the two classic \hmxbs, Vela X-1 and Cygnus X-1, for which we provide, for illustrative purposes, the corresponding scales in Table\,\ref{tab:norm}. However, our aim is not to limit our study to these two systems specifically but rather to explore a set of realistic configurations which are thought to cover most of the known \sgxs, including Vela X-1 and Cygnus X-1. From now on, unless specified otherwise, the equations are given in their dimensionless form.

\begin{table}[]
\caption{Dimensionless parameters of the model and the values we consider for each of them. Orbital periods were chosen to match, after normalization, the values in Vela X-1 and Cygnus X-1 respectively but they could also correspond to other systems. Normalization units can be found in Table\,\ref{tab:norm}.}
\label{tab:params}
\begin{tabular}{ll}
\hline\hline
$m_{cl}/m_0$ & 1$\cdot$10$^{-5}$, 4$\cdot$10$^{-7}$\\ 
$R_{cl,2}/R_*$ & 0.005, 0.01, 0.02, 0.04, 0.08\\ 
$\beta$ & 0.5, 2.0\\ 
$a/R_*$ & 1.6, 2, 3\\ 
$i$ & \ang{0}, \ang{25}, \ang{45}, \ang{65}, \ang{90}\\ 
$P/t_0$ & 24, 85\\ 
\hline
\end{tabular}
\end{table}

\begin{table}[]
\caption{Scaling of the model and typical physical values for two representative \sgxs, Vela X-1 \citep[from][]{ElMellah2018} and Cygnus X-1 \citep{Herrero1995,Orosz2011}. In the upper part, the three fundamental normalization units are given. In the lower part, time, mass and column density normalization units are provided for the two same systems (we assume a mean molecular weight of one proton mass $m_p$).}
\label{tab:norm}
\begin{tabular}{lll}
\hline\hline 
\hmxbs & Vela X-1  & Cygnus X-1 \\ 
\hline 
$R_*$ &  28 $R_{\odot}$        &  17 $R_{\odot}$        \\
$v_{\infty}$ &  600  km$/$s        &  2\,100  km$/$s        \\
$\dot{M}$ & 6.3$\cdot$10$^{-7}$\msun$\cdot$yr$^{-1}$          &  3.0$\cdot$10$^{-6}$\msun$\cdot$yr$^{-1}$         \\ \hline
$t_0=R_*/v_{\infty}$ & 33 ks & 5.7 ks         \\
$m_0=\dot{M}t_0$ & $1.3\cdot 10^{24}$g & $1.1\cdot 10^{24}$g  \\
$N_{H,0}=m_0/(m_p R_*^2)$ & $2.0\cdot 10^{23}$cm$^{-2}$        & $4.6\cdot 10^{23}$cm$^{-2}$ \\
\hline
\end{tabular}
\end{table}

\subsection{Numerics}
\label{sec:XXX}

\subsubsection{Clumps}
\label{sec:XXX}

We work in the observer frame centered on the donor star and neglect stellar rotation and wobbling around the center of mass. The latter assumptions are expected to have little impact on the column density variability since the wind essentially flows away from the star, with terminal wind speeds typically larger than the orbital speed by a factor of a few. The position of the clumps is advanced from a lower spherical boundary of radius $r_m=1.2$ to an upper spherical boundary of radius $r_M$. Between $r=1$ and $r=r_m$, the wind is assumed to be smooth, to represent the finite distance from the photosphere beyond which significant clumping is expected \red{\citep[see Fig. 1 for theoretical estimates, and for example][for observational constraints]{Puls2006}}. Clump emission is isotropic in the sense that clumps are injected at the inner boundary $r=r_m$ with an equal probability per solid angle. At each time step, clumps beyond $r=r_M$ are deleted while clumps are created at the basis at a rate $\dot{\mathcal{N}}=1/m_{cl}$. We can estimate a priori the number $\mathcal{N}$ of clumps in the simulation space by writing:
\begin{equation}
    \mathcal{N}=\int_{r_m}^{r_M} n 4\pi r^2 dr=\frac{1}{m_{cl}}\int_{r_m}^{r_M}\frac{dr}{v}.
\end{equation}
A preliminary integration phase of duration $\mathcal{N} m_{cl}$ guarantees that when we start computing column densities, the whole simulation space is populated with clumps, from the lower to the upper radial boundaries. The position of each clump is advanced using a 4$^{th}$ order Runge-Kutta integration method, with a time step $\Delta t$ equal to $P/(3\cdot10^4)$. This time step can be compared to the minimum of the duration for a clump to pass in front of a static point source:
\begin{equation}
    \label{eq:crossing_time}
    \Delta t=\text{min}\left[2R_{cl}(r)/v(r)\right],
\end{equation}
which is found, for both expansion laws, at $r=\beta+1$. Even in the least favorable configuration, this characteristic time scale is still hundreds of times larger than $\Delta t$, which ensures that we resolve the impact of all clumps on the column density, even the smallest ones with grazing impact parameters with respect to the \los. For realistic parameters and $r_M=30a$ (see next section), the total number of clumps can reach up to 10$^9$ which need to be integrated over $3\cdot10^5$ of time steps in order to get the evolution of the column density over 10 orbital periods, our full integration time. In order to reduce the number of clumps and speed up the computation, we retain only the clumps susceptible to intercept the \los and discard the others. As visible in Figure\,\ref{fig:big_picture}, it means that the number of clumps to consider is much smaller, in particular when the orbit is close to edge-on.

\subsubsection{Maximum distance of integration}
\label{sec:XXX}

\begin{figure}
\centering
\includegraphics[width=0.99\columnwidth]{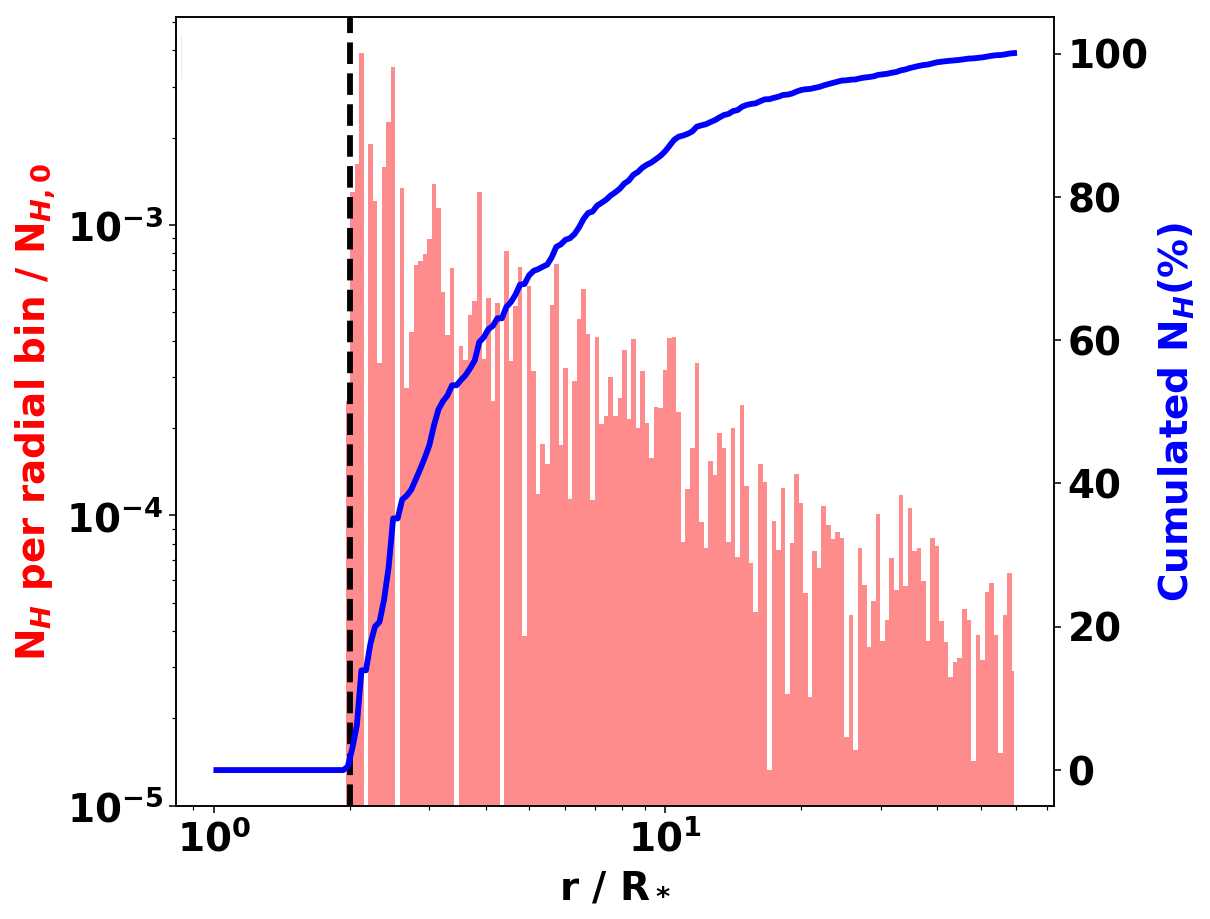}
\caption{Column density, per logarithmic radial bin (red, logarithmic axis on the left) and cumulative (blue, linear axis on the right), as a function of the distance to the stellar center. The position of the compact object is indicated with a dashed black line. The total column density up to 30 orbital separations is $\sim$97\% of the column density computed for a smooth wind integrated up to thousands of stellar radii. The dimensionless parameters are $m_{cl}=10^{-5}$, $R_{cl}=0.08$, $a=2$, $P=80$ and $i=45^{\circ}$.}
\label{fig:NH_shells}
\end{figure}

Every time step, we cast a ray from the X-ray point source to the observer and sum up the contribution of each segment which intercepts a clump to obtain the total column density at this time step. To determine at which distance $r_M$ we should stop the integration of the column density along the \los, we quantified the contribution of each logarithmic radial bin between the compact object and large distances. We worked at inferior conjunction ($\phi=0.5$) such as these results are a fortiori valid when the X-ray source is buried deeper into the wind. In Figure\,\ref{fig:NH_shells} is represented the cumulated column density in blue, and the contribution of each bin in red. As we move away from the accretor (located at $r=2$ in this example), the column density per logarithmic radial bin decreases quickly: 10 orbital separations away from the stellar center, the column density integrated over a certain interval is $\sim$50 times lower than the one integrated over a range 10 times shorter but around $r=a$. This common trend for all the sets of parameters we considered drove us into setting the outer edge of the simulation domain at $r_M=30a$, a distance at which the column density has essentially converged.

\section{Analytical predictions}
\label{sec:analytics}

\subsection{Smooth wind scenario}
\label{sec:analytics_smooth}

In the case of a smooth (\ie homogeneous, without clumps) and isotropic wind with a radial velocity profile which follows a $\beta$-law, the \red{normalized} column density profile along the orbit of the accretor can be written as: 
\begin{equation}
\label{eq:NH_smooth}
    N_H(\phi)=\frac{1}{4\pi}\int_{z(\phi)}^{\infty}\frac{dz}{r^2\left(1-1/r \right )^{\beta}},
\end{equation}
where $r$ (function of $z$ and of the orbital elements) and $z$ are respectively the distance to the stellar center and the coordinate along the \los joining the X-ray point source to the observer, both normalized with the stellar radius $R_*$, and $\phi$ is the orbital phase (from 0 to 1). This profile depends on three shape parameters and one scale. The three shape parameters are the exponent $\beta$ of the velocity profile and the orbital elements. For a circular orbit, the latter are the inclination angle $i$ of the orbital angular momentum with respect to the \los and $a$, the orbital separation divided by the stellar radius. They both play a role to determine the inferior bound of the integral, $z(\phi)$, which locates the compact object along the \los, and to convert $r$ into $z$. The integral can then be performed numerically to compute any column density profile in a smooth and isotropic wind. In eclipsing \hmxbs, the median $N_H$ profile at egress is the most constraining due to its sharp decrease. The fit of the measured column density profile with equation\,\eqref{eq:NH_smooth} also grants access to the mass loss rate via the column density scale $N_{H,0}$, provided we can estimate the terminal wind speed and the stellar radius (see Table\,\ref{tab:norm} for scalings). 

To illustrate how $a$, $i$, $\beta$ and $N_{H_0}$ can be obtained based on this fitting procedure, we focus on an actual \hmxb. The classic \sgx 2S0114$+$650 (also known as LSI$+$65 010) is an excellent candidate to apply the model presented in this paper, thanks to its grazing orbit. \cite{Sanjurjo-Ferrrin2017} provide a ratio $a/R_*$ of 1.34-1.65. Given the high mass ratio \citep{Reig1996}, the absence of accretion disk \citep{Koenigsberger2003} and the moderate X-ray luminosity \citep[$1.4\cdot10^{36}\text{erg}\cdot\text{s}^{-1}$,][]{Sanjurjo-Ferrrin2017}, we can safely assume that the star does not fill its Roche lobe. It brings us to the conclusion that the upper limit of this range is to be preferred for the ratio $a/R_*$. \cite{Pradhan2015} showed that it was not eclipsing by measuring a finite column density at superior conjunction. However, the sharp decrease of the column density following superior conjunction indicates that the \los probes deep layers of the wind, near the photosphere. Consequently, the orbital inclination has to be between $\ang{40}$ and $\ang{50}$ approximately, which is also consistent with results obtained from radial velocity measures of the donor star \citep{Crampton1985,Farrell2008b,Koenigsberger2006}. We fit the $N_H$ data between orbital phases 0 and 0.1 reported in \cite{Pradhan2015} with the smooth wind profile given by equation\,\eqref{eq:NH_smooth}. We could find a reasonable match for the following parameters: $a=1.6$, $i=\ang{50}$, $\beta=1$ and $N_{H,0}=1.2\cdot 10^{23}$cm$^{-2}$. \cite{Reig1996} estimated the terminal wind speed to be 1\,200 km$/$s, such that if the stellar radius is $R_*=25$R$_{\odot}$, we deduce from $N_{H,0}$ a stellar mass loss rate $\dot{M}\sim 7\cdot 10^{-7}$\msun yr$^{-1}$, very similar to the inferred mass loss rate of the donor star in Vela X-1 \citep{Gimenez-Garcia2016}. Spectral characterization in the optical and UV waveranges of the stellar atmosphere with non local thermodynamical equilibrium analysis can bring additional constrains to break the degeneracy and disentangle between the different parameters which come into play in the value of $N_{H,0}$ \citep{Hainich2020}.

Due to the conservation of mass, the averaged column density should remain the same whether the mass is gathered in clumps or spread in a smooth wind: the properties of a smooth wind are expected to match the time-averaged properties of a clumpy wind where clumps are realistically small and numerous, a conjecture which will be validated in Section\,\ref{sec:peaktopeak}. The micro-structure of the wind will only manifest through the time variability of the column density.

\subsection{Standard deviation}
\label{sec:analytics_std}

Following the arguments by \cite{Owocki2018}, we may write the variance of the column density between a source located at $z_0$ along a \los and an observer at $z = \infty$ as, in physical units and for a mean molecular weight of one proton mass $m_p$:
\begin{equation} 
    \label{eq:delta_N}
    \delta N_H^2 = \int_{z(\phi)}^{\infty} \left(\frac{\rho}{m_p}\right)^2 h\,dz  ,
\end{equation} 
where $h$ is again the porosity length and \red{$\rho = \langle \rho \rangle$ is the mean mass density
of the isotropic wind.}  
Using the continuity equation in combination with equation\,\eqref{eq:porosity_length}, we obtain the dimensionless standard deviation $\delta N_H$ in column density around the value set by the smooth and isotropic wind as a function of the orbital phase: 
\begin{equation}
    \label{eq:delta_NH_general}
    \delta N_H =\frac{3}{32\pi}\left(m_{cl}\int_{z(\phi)}^{\infty} \frac{dz}{R_{cl}^2 r^2 v} \right)^{1/2}.
\end{equation}
The expansion law for the clumps can then be reinjected in the formula above. For example, for linearly expanding clumps, we then have, if the velocity profile is given by the $\beta$-law \eqref{eq:beta_law}:
\begin{equation}
    \delta N_H=\frac{3}{8\pi}\left(\frac{m_{cl}}{R_{cl,2}^2}\int_{z(\phi)}^{\infty} \frac{dz}{r^4 \left(1-1/r \right)^{\beta}} \right)^{1/2}.
\end{equation}
This shows explicitly how, for a given orbit and wind acceleration, the standard deviation of the column density depends on the ratio $\sqrt{m_{cl}}/R_{cl,2}$ for any power-law clump radius expansion.

We performed simulations with both clump radius expansion laws for the 300 simulations with high-mass clumps ($m_0=10^{-5}$), less computationally-demanding due to the lower total number of clumps in the simulation space. The results were similar enough that all the conclusions we draw in the next sections are equally valid for both laws. It was to be expected because in addition to the expression\,\eqref{eq:delta_NH_general} above, Figure\,\ref{fig:porosity} indicates that clump radii and porosity lengths within 3$R_*$, in the region where the clumps contribute the most to the final column density, are alike. Hereafter, results are presented for the smoothly expanding law\,\eqref{eq:lorenzo_ext}.

\section{Results}
\label{sec:results}

In this section, we focus on the results of our simulation, in particular the time-evolution of column density (Sec.~\ref{sec:NH_lcs}), peak to peak variability (Sec.~\ref{sec:peaktopeak}), and coherence time scales (Sec.~\ref{sec:coherence}. We compare to theoretical predictions and relate the so derived properties of absorption variability to the underlying properties of the clumpy winds. The observational implications and in particular the challenges and prospects are discussed later, in Sec.~\ref{sec:limitations}.

\subsection{Time evolution of the column density}
\label{sec:NH_lcs}

\begin{figure*}
\centering
\includegraphics[width=1.9\columnwidth]{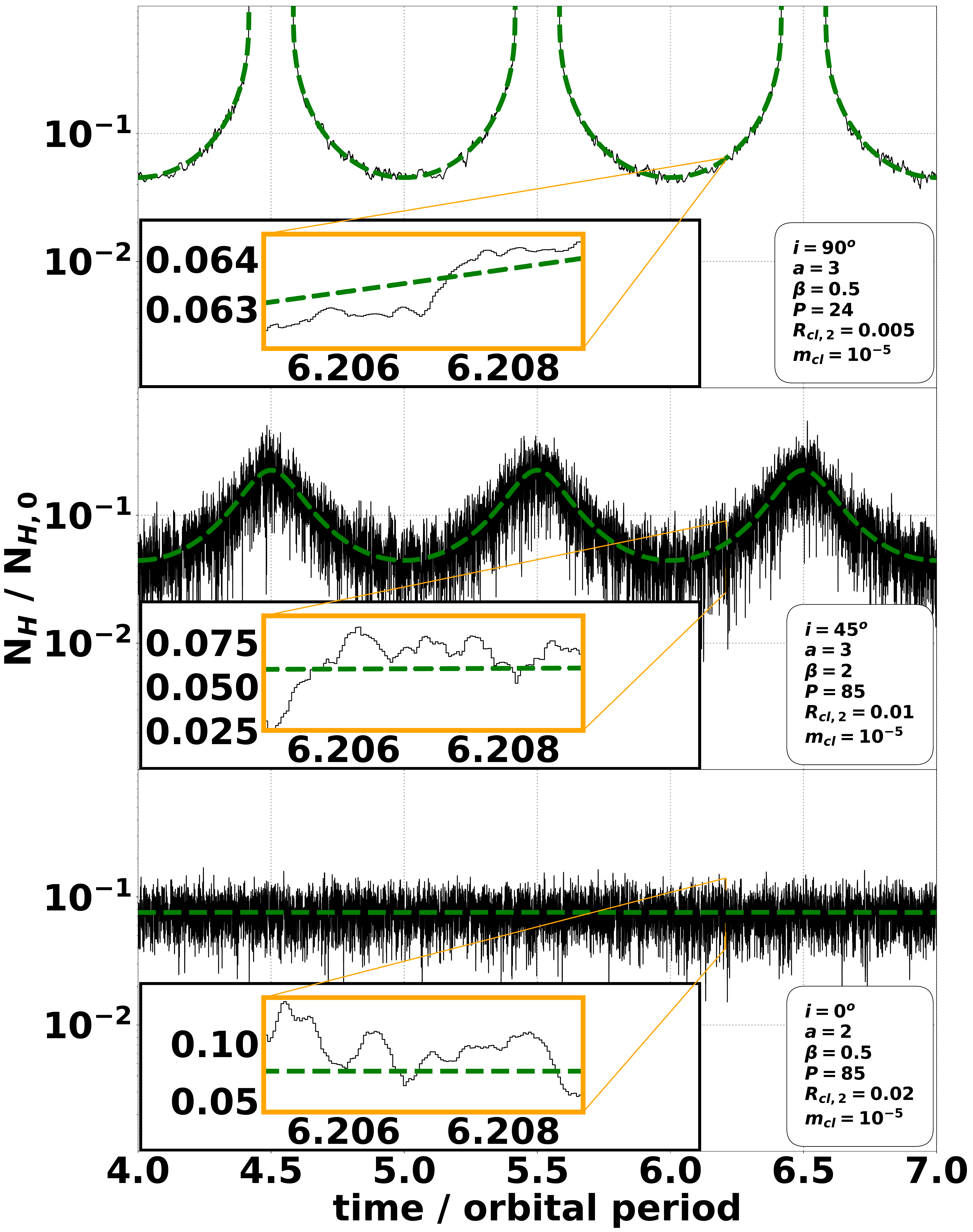}
\caption{Three different evolutions of the column density as a function of time over several orbits for $i=\ang{90}$ (top panel), $i=\ang{45}$ (middle panel) and $i=\ang{0}$ (bottom panel) - and different clump radii, orbital separations, periods and $\beta$ exponent. The dashed green line is the corresponding smooth wind column density profile. The top profile displays eclipses due to the high inclination. The insets show zoomed in portions of the curves where the numerical time sampling of the column density, $3\cdot 10^4$ times per orbital period, is visible. The origin of time is set at inferior conjunction (\ie $t=0$ at $\phi=0.5$).}
\label{fig:NH_lcs}
\end{figure*}

The main output of each simulation is a time series of the column density. This data is represented for three simulations in Figure\,\ref{fig:NH_lcs} over a few orbital periods. As the compact object orbits the donor star, the amount of material integrated along the \los changes because of two effects: the motion of the X-ray point source around the star and the clumps crossing the \los. The latter induces a variability on short time scales whose statistical properties relate to the micro-structure of the wind and are discussed in more detail in Sections\,\ref{sec:peaktopeak} and \ref{sec:coherence}. The orbital motion, on the other hand, causes a smooth periodic modulation of the column density, more contrasted for higher inclination and orbital separation, and which vanishes for face-on configurations where the distance between the accretor and the observer remains constant (see bottom panel in Figure\,\ref{fig:NH_lcs}). At high inclination, eclipses can occur during which the column density can not be measured (see gaps in the top panel in Figure\,\ref{fig:NH_lcs}, for a \ang{90} inclination). The duration $d$ of an eclipse with respect to the orbital period $P$ relates to the ratio $a/R_*$ via:
\begin{equation}
\label{eq:eclipse_duration}
    \frac{d}{P}=\frac{1}{\pi}\arcsin\left(\frac{R_*\sin i}{a}\right)\sim \frac{1}{\pi}\frac{R_*\sin i}{a},
\end{equation}
where the approximation is valid within 10\% percent for any inclination provided $a>1.6R_*$, a condition usually met in \hmxbs \citep[\eg][]{Falanga2015}. In Figure\,\ref{fig:NH_lcs}, we represented the smooth wind solution obtained with equation\,\eqref{eq:NH_smooth} with a dashed green line. The measured column densities oscillate around this profile. Because we only cover the case of circular orbits in this paper, the smooth wind as a function of the orbital phase $\phi$ is symmetric with respect to the axis $\phi=0$ and $\phi=0.5$ (superior and inferior conjunction respectively). We highlighted the short term variability of the column density with inserts showing zoomed-in versions of the time series. Each insert contains approximately 200 data points which are visible and indicate that the impact of the clumps is resolved up to very short time scales. The total column density consists of cumulative contribution of many clumps with different radii and densities, which indicates that accurate knowledge on the clumpiness of the stellar wind can be gained from the time variability of the column density.

\subsection{Column density peak-to-peak variability}
\label{sec:peaktopeak}

\subsubsection{Variability amplitude as a function of the orbital phase}
\label{sec:XXX}

\begin{figure}
\centering
\includegraphics[width=\columnwidth]{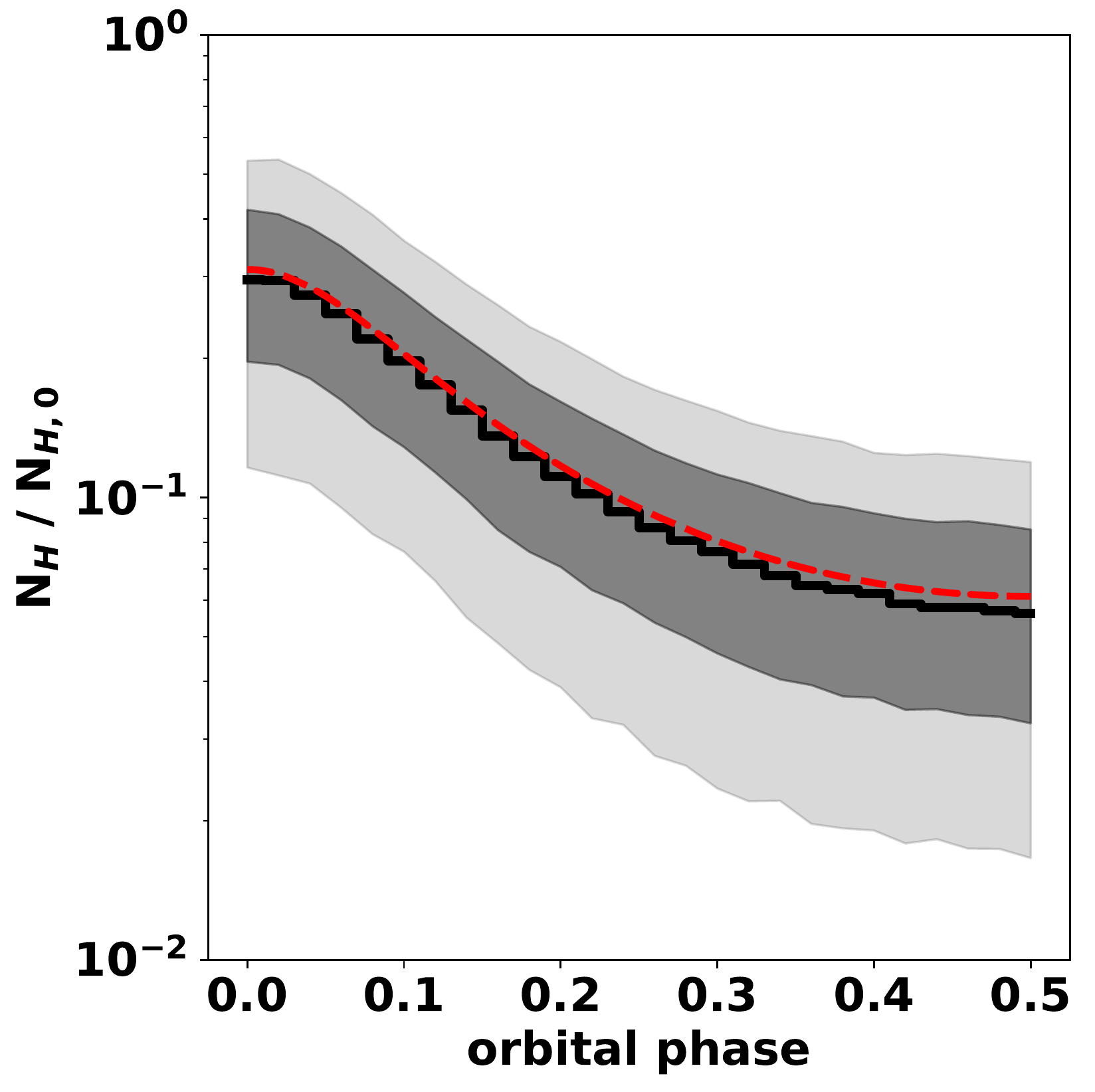}
\caption{Column density from superior conjunction ($\phi=0$) to inferior conjunction ($\phi=0.5$) for an orbital inclination $i=\ang{45}$. In black is represented the median over 10 orbits and the red dashed line is the smooth wind solution from equation\,\ref{eq:NH_smooth}. The dark (resp. light) grey shaded region is bounded by the 20$^{th}$ to 80$^{th}$ percentiles (resp. the 5$^{th}$ to 95$^{th}$ percentiles). The parameters are $m_{cl}=10^{-5}$, $R_{cl}=0.01$, $\beta=0.5$ and $a=2$.}
\label{fig:landscape}
\end{figure}

In Figure\,\ref{fig:NH_lcs}, the spread of the column density around its smooth wind value changes from a simulation to another. In this section, we determine the main parameters which set the measured standard deviation but beforehand, we represent the computed column density as a function of the orbital phase in Figure\,\ref{fig:landscape} for a fiducial set of parameters. We computed the column density during each simulation at 50 locations equally spaced on the orbit simultaneously every 100 time steps such as each orbital phase bin in Figure\,\ref{fig:landscape} contains 6\,000 values of $N_H$ (after folding the orbit from $\phi=0$ to $\phi=0.5$). For each bin, we compute the median (black steps) and the standard deviation. The black shaded region around the median shows the values between the 20$^{th}$ and the 80$^{th}$ percentiles while the grey shaded region is delimited by the 5$^{th}$ and the 95$^{th}$ percentiles. Simultaneous computation of $N_H$ along the orbit enables us to get a much better statistics to compute the standard deviation at each orbital phase than using the time series presented in Section\,\ref{sec:NH_lcs}. It does not introduce biases due to time coherence of the signal associated to clumps if the simulation duration (ten orbital periods) is high compared to the coherence time scale, an assumption we validate a posteriori (see Section\,\ref{sec:coherence}). These results indicate that except for very large ($R_{cl}=0.08$) and sparse clumps ($m_{cl}=10^{-5}$), the median $N_H$ profile is representative of the smooth wind profile with the same parameters but where clumps are replaced with a smooth wind of same mass loss rate. It means that when a column density is measured from observations around a certain orbital phase (\eg during a 10 ks observation, which would correspond to an orbital phase width on the order of 1$/$50 in Cygnus X-1), it can be compared to Figure\,\ref{fig:landscape} for a set of parameters believed to match the observed system. If the column density at this very same orbital phase is repeatedly measured over different orbits, the histogram of values should reproduce a vertical slice in Figure\,\ref{fig:landscape}, with the median corresponding to the value for a smooth wind of same mass loss rate.


\subsubsection{Comparison to analytical predictions}
\label{sec:comp_to_analytics}

\begin{figure}
\centering
\includegraphics[width=0.99\columnwidth]{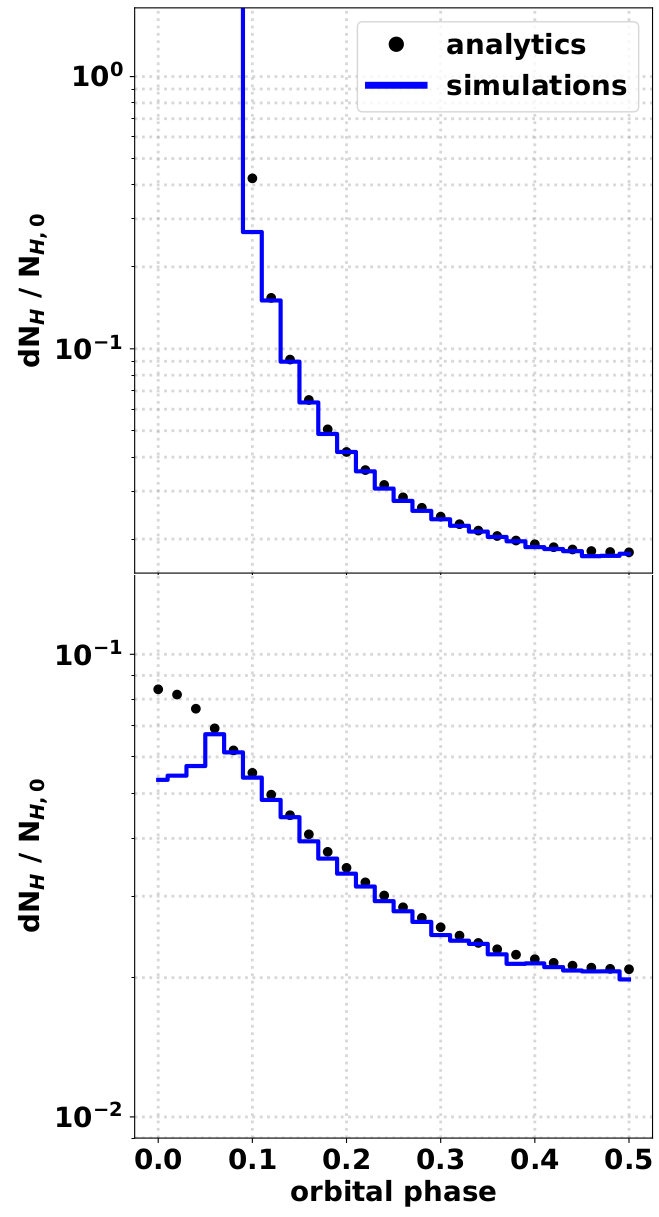}
\caption{Standard deviation of the column density as a function of the orbital phase computed numerically (in blue) and analytically (black dots) for simulations with $m_{cl}=10^{-5}$, $R_{cl}=0.02$. In the top panel, $\beta=2$, $a=2$ and $i=\ang{90}$ and in the bottom panel, $\beta=0.5$, $a=1.6$ and $i=\ang{45}$. The discrepancy at low orbital phase is partly indicative of the region near the stellar photosphere where the wind in the numerical model is assumed to not be clumpy yet. }
\label{fig:dNH_ph}
\end{figure}

For each simulation, we compared the profiles of the standard deviation as a function of the orbital phase to the predicted profiles derived in Section\,\ref{sec:analytics_std} (see Figure\,\ref{fig:dNH_ph}). Apart near superior conjunction in a few configurations, they match within a few percent. The discrepancies sometimes observed near superior conjunctions can be fully attributed to two effects. 
First, when the system is eclipsing, the orbital bin which contains ingress or egress underestimates the variability because it averages the standard deviation over a finite orbital phase interval which partly includes the eclipse, during which there is no variability. Second and more importantly, when the \los intercepts the region $r\in [1;1.2]$ around the donor star where we assumed that the wind is smooth, the standard deviation is also lowered with respect to the prediction of the analytical model where the wind is assumed to be clumpy all the way down to the stellar photosphere. 

Observational hints in favor of clumpiness near the star exist \citep{Puls2006,Cohen2011,Torrejon2015} but our model provides a way to quantify this: if the standard deviation at superior conjunction measured in \hmxbs where the orbit is grazing is lower than what predicted from fitting the standard deviation profile far from superior conjunction with the formula\,\eqref{eq:delta_N}, then it would support the scenario of a smooth wind base. Monitoring the $\delta N_H$ profile in highly inclined systems might thus enable us to constrain the onset of the clump-forming region. For instance, the bottom panel in Figure\,\ref{fig:dNH_ph} corresponds to a configuration where, although the system is not eclipsing (since at superior conjunction, when $\phi=0$, the projected distance to the star is $a\sin i > 1$), the \los to the X-ray source probes the deep smooth layers near the photosphere (since at superior conjunction, $a\sin i < 1.2$). We see that the standard deviation is lower than predicted by the analytic model (which assumes that the wind is clumped all the way down to the stellar surface) up to an orbital phase of $\sim$0.07, when the \los no longer intersects the smooth wind region. In \sgxs where the orbital inclination and separation are such that the \los probes the smooth wind region near the star, the location of the maximum in the $\delta N_H$ profile is thus a direct measure of the distance to the stellar photosphere where clumping occurs.

\begin{figure*}
\begin{center}
\includegraphics[width=16cm]{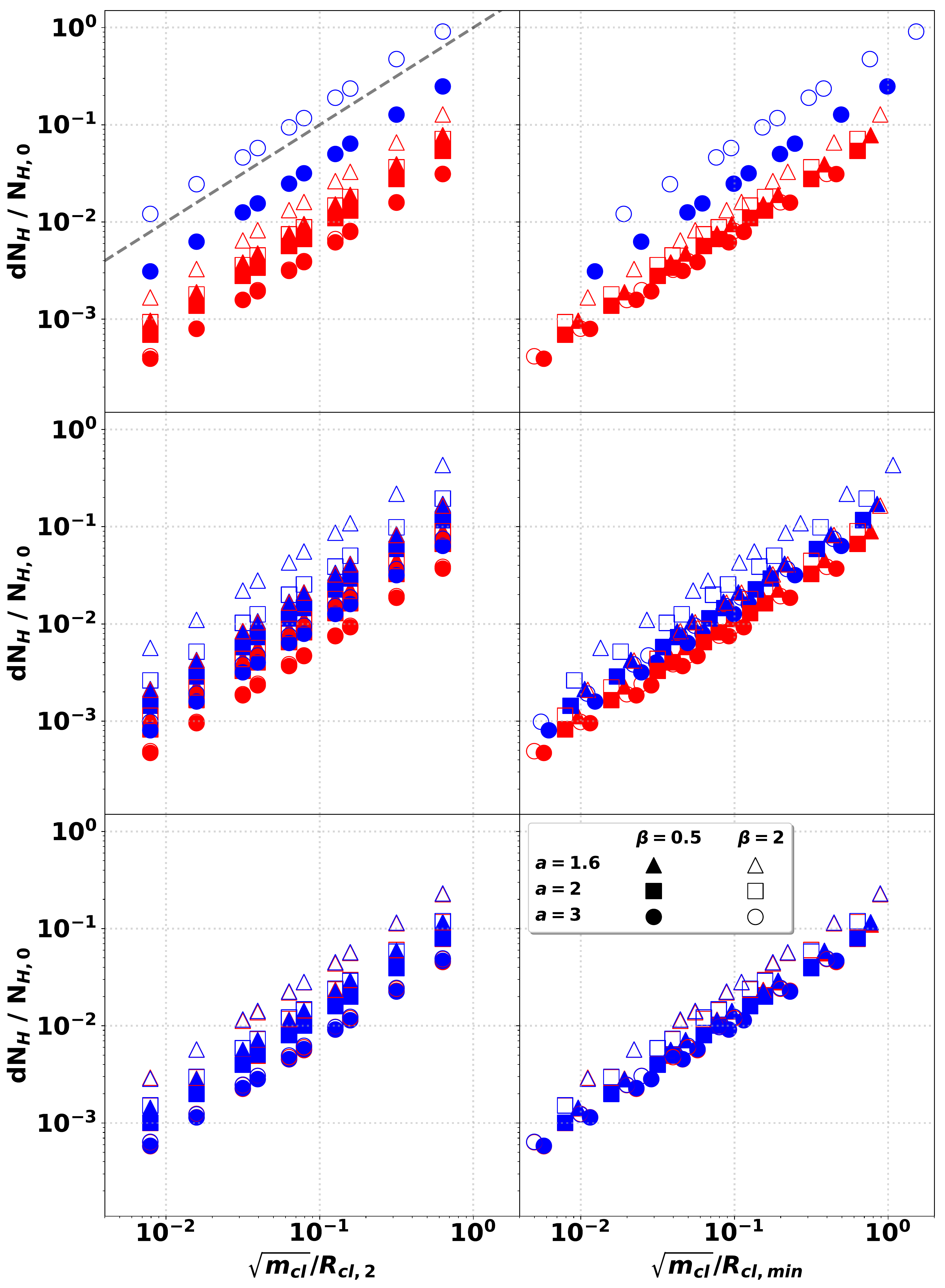} 
\end{center}
\caption{From top to bottom row, the orbital inclination is \ang{65}, \ang{25} and \ang{0}. (left column) Standard deviation of the column density at inferior (resp. superior) conjunction in red (resp.blue) as a function of the ratio $\sqrt{m_{cl}}/R_{cl,2}$. Each couple $(a,\beta)$ leads to a different proportionality constant between $\delta N_{H}$ and $\sqrt{m_{cl}}/R_{cl,2}$, hence the large dispersion. The dashed line in the top left panel indicates the $y=x$ line. (right column). Same as before but $R_{cl,2}$ has been replaced by the clump radius at the point along the \los between the X-ray source and the observer the closest from the donor star. The dispersion is reduced, especially at inferior conjunction.}
\label{fig:sigma}
\end{figure*}

\subsubsection{Dependence on parameters}
\label{sec:dNH_params}

Finally, we compare the standard deviation of the column density at inferior and superior conjunctions, $\delta N_{H,inf}$ and $\delta N_{H,sup}$, for different parameters. The dependency of $\delta N_{H}$ on $\sqrt{m_{cl}}/R_{cl}$ in equation\,\ref{eq:delta_N} invited us to represent $\delta N_{H,inf}$ and $\delta N_{H,sup}$ as a function of $\sqrt{m_{cl}}/R_{cl,2}$ in the left column in Figure\,\ref{fig:sigma} for an orbital inclination of \ang{65}, \ang{25} and \ang{0} (top, middle and bottom rows). For eclipsing \hmxbs, the standard deviation at superior conjunction is not defined and thus not represented. We vary the orbital separation (different markers) and differentiate quickly and slowly accelerating winds (filled and empty markers respectively). In blue (resp. red) are shown the values of $\delta N_{H,sup}$ (resp. $\delta N_{H,inf}$). These results do not depend on the orbital period, which was expected since the orbital period is so much longer than the time scale at which $N_H$ varies. All other parameters being fixed, we see that the relation $\delta N_{H}\propto\sqrt{m_{cl}}/R_{cl,2}$ whose slope is indicated in the first panel with a black dashed line is verified for each set of ten points (five different values of $R_{cl,2}$ and two values of $m_{cl}$). The ratio between the standard deviation at superior and inferior conjunctions is hardly sensitive to the orbital separation. As the orbital separation increases, the difference between slowly and quickly accelerating winds vanish at inferior conjunction: while the wind speed at $r=1.6$ is more than four times higher for $\beta=0.5$ than for $\beta=2$, this ratio drops to less than 2 at $r=3$. The difference between $\beta=0.5$ and $\beta=2$ remains significant at superior conjunction though, even for $r=3$, since the \los probes regions $r<a$ enclosed within the orbit. A higher $\beta$ leads to a higher standard deviation due to the higher number of clumps per unit volume at a given distance from the star (see equation\,\eqref{eq:mdot_ndot}). The standard deviation at superior conjunction is always higher than at inferior conjunction, except for face-on orbits where they coincide: at superior conjunction, the \los intercepts a region closer from the star than at inferior conjunction, where the clumps are small and packed, which increases the standard deviation.

Given the dominating contribution of the densest clumps pinpointed in Figure\,\ref{fig:NH_shells}, we changed the clump radius used in the x-axis of the plots in Figure\,\ref{fig:sigma} so as it corresponds to the smallest clump radius encountered along the \los, $R_{cl,min}$, that is the clump radius at the point along the \los the closest from the donor star (right column in Figure\,\ref{fig:sigma}). At inferior conjunction, this point is simply the position of the compact object but at superior conjunction, it is a point located deep in the wind if the inclination is significantly high, such as the relevant clump radius becomes, for smoothly expanding clumps:
\begin{equation}
  R_{cl,min}(\phi=0)=R_{cl,2}\left(\frac{a\cos i}{2}\right)^{2/3}\left(2-\frac{2}{a\cos i}\right)^{\beta/3}.
\end{equation}
This correction enables us to get a much lower dispersion of the points along the $\delta N_{H}\propto\sqrt{m_{cl}}/R_{cl,min}$ axis and proves that the standard deviation is an excellent tracer of the ratio $\sqrt{m_{cl}}/R_{cl,min}$. Once the parameters $a$, $i$, $\beta$ and $N_{H,0}$ have been constrained thanks to the median column density profile as a function of the orbital phase, a precise measure of the standard deviation grants us access to the ratio $\sqrt{m_{cl}}/R_{cl,min}$. The accuracy of this obtained ratio is maximal if $\delta N_H$ is measured at inferior conjunction, where the dispersion is the lowest. The latter effect can be explained by the fact that at inferior conjunction, the \los is best aligned with the radial direction from the donor star (with exact alignment for an edge-on system) so the column density is mostly localized near the accretor and the contributions quickly drop as we go toward the observer. Another advantage of working at inferior conjunction is to obtain an accurate median profile $N_H\left(\phi\right)$ with fewer observations. The spread which subsists in the right column in Figure\,\ref{fig:sigma} is due to the minor but non negligible contribution of clumps with radii $R_{cl}>R_{cl,min}$ on the standard deviation of the column density.

These results illustrate how the standard deviation on column density can be exploited to constrain the ratio $\sqrt{m_{cl}}/R_{cl,min}$, 
\red{or alternatively the porosity length since $\sqrt{m_{cl}}/R_{cl} \propto \sqrt{h}$. In order to disentangle the clump mass and radius,} 
it is necessary to analyze another output of our simulations, the coherence time scale.






\subsection{Coherence time scale}
\label{sec:coherence}

\subsubsection{Autocorrelation function}
\label{sec:XXX}

\begin{figure}
\centering
\includegraphics[width=\columnwidth]{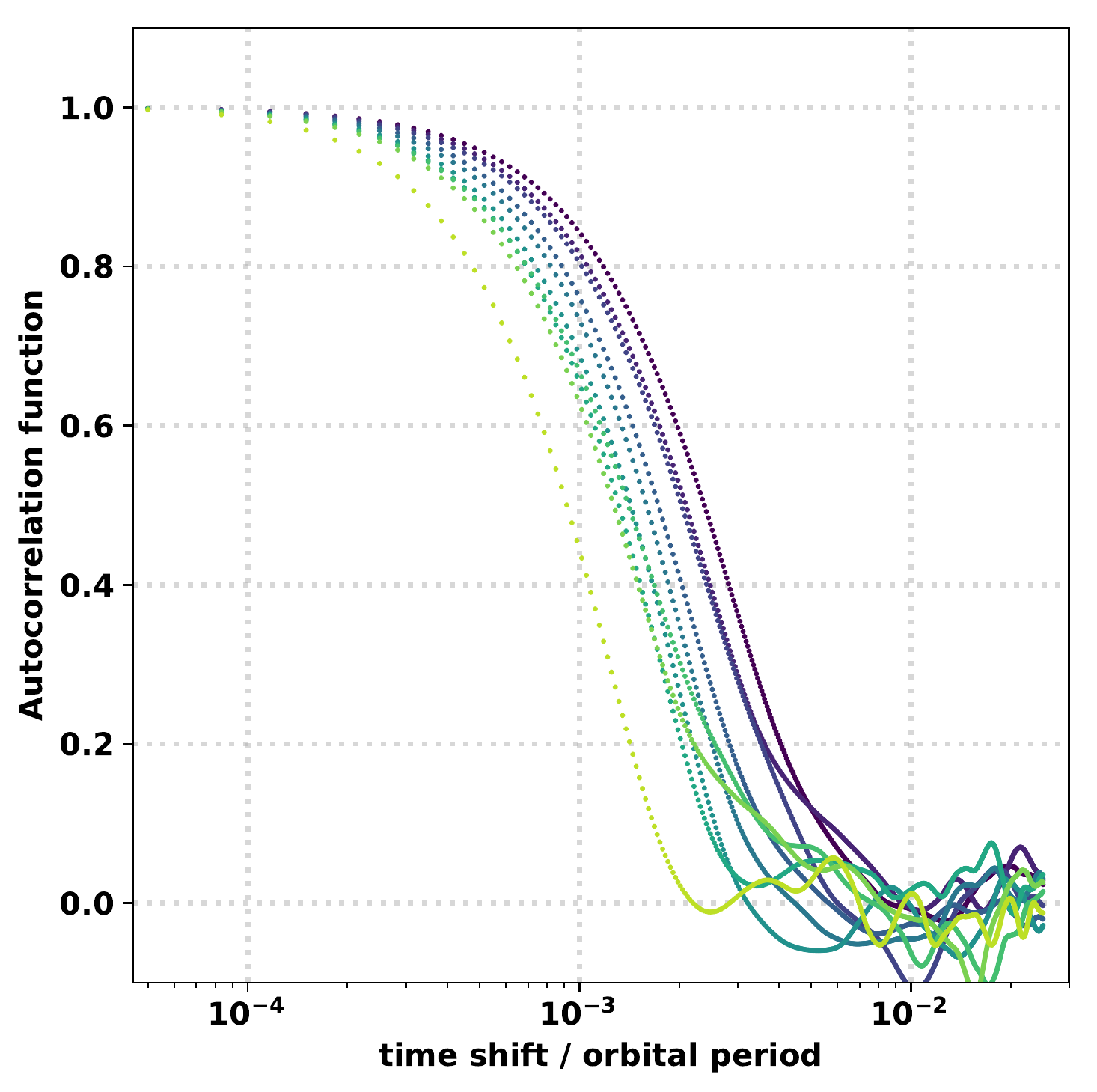}
\caption{Autocorrelation function of the column density as a function of the time shift. From purple to green and yellow, we represent the autocorrelation functions of the data at different orbital phases, from inferior conjunction (dark purple) to superior conjunction (yellow). Simulation with $m_{cl}=4\cdot10^{-7}$, $R_{cl}=0.04$, $\beta=2$, $a=3$, $i=\ang{65}$ and $P=85$.}
\label{fig:CCF}
\end{figure} 

In the previous section, we extracted information on the characteristic standard deviation of the column density at each orbital phase for each set of parameter, but we did not exploit the information on the time evolution of the column density. 
The time series contain additional information, the coherence time scale, which is the characteristic duration over which the changes in column density are smooth and, on average, predictable. If observations of \hmxbs are carried out over intervals lasting a significant fraction of the orbital period, the column density as a function of time can, in theory, be extracted on time scales shorter than this coherence time scale (see Sec.~\ref{sec:obs_chal} for example estimate of necessary time resolution).

In order to determine the coherence time scales in each simulation, we first compute the relative difference between the time series presented in Section\,\ref{sec:NH_lcs} and the smooth wind solution \eqref{eq:NH_smooth}, and subtract the median of the obtained signal afterwhile. Since we expect the coherence time scale to vary with the orbital phase, we must treat separately the different orbital phases to avoid smearing out the coherence time scale. \blue{We cut the time series in twenty equal orbital phase intervals. In spite of the motion of the X-ray source along its orbit, we can still use the symmetry with respect to $\phi=0.5$ and gather the data at $\phi=0.25$ and $\phi=0.75$ for instance, since both are expected to have the same coherence time scale. For each simulation, we are left with ten orbital phase intervals of width 0.05, between $\phi=0$ and $\phi=0.5$, each containing twenty time series of the column density of duration $P/20$. The autocorrelation function of each of this individual time series is then computed and the twenty autocorrelation functions of each orbital phase bin are averaged to obtain a representative autocorrelation function for each of the ten intervals\footnote{We note that while we employ the autocorrelation function here, an alternative but mathematically equivalent approach using power density spectra to assess the characteristic time scales of variability can be applied.}. With this method, we can capture coherence time scales ranging from $P/(3\cdot 10^4)$ (the time resolution) up to $P/40$.} As an illustration, the ten autocorrelation functions of a fiducial simulation are represented in Figure\,\ref{fig:CCF}, from inferior conjunction (dark purple) to superior conjunction (yellow).

The autocorrelation function is maximal in zero since the time series is fully correlated to its zero time shifted counterpart. As the shift increases, the correlation decreases until the time shift is so large that the two functions are no longer correlated at all. The coherence time scale corresponds to the time beyond which the signal changed enough to consider that it lost the memory of its previous values. We measure it by taking the time shift when the autocorrelation has been halved, $\tau$. We evaluate the uncertainty on the obtained value of $\tau$ by looking at the face-on configurations, where $\tau$ should not depend on the orbital phase: it remains indeed constant, and varies by at much 5\%.

The coherence time scale measured at inferior conjunction is typically higher than the one at superior conjunction. The \los probes deeper wind layers at superior conjunction where clumps are smaller but also slower, such that a shorter coherence time scale is not obvious without accounting for the orbital speed, which turns out to be larger than the speed of the clumps near the stellar photosphere.


\subsubsection{Dependence on parameters}
\label{sec:XXX}

\begin{figure}
\centering
\includegraphics[width=\columnwidth]{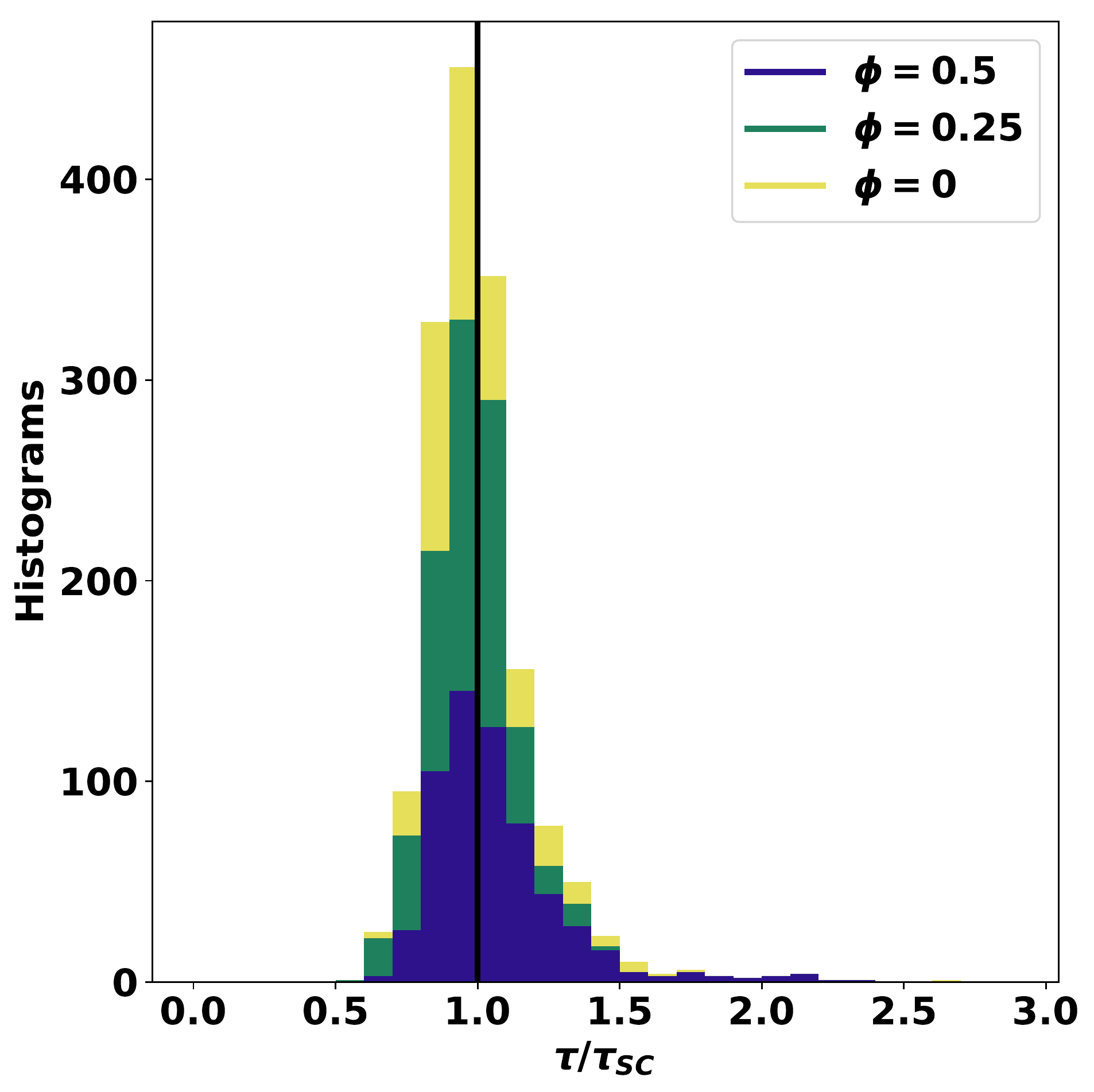}
\caption{Stacked histograms of the ratios of the measured coherence time scale $\tau$ compared to the clump self-crossing time $\tau_{SC}$ for all 600 simulations at inferior conjunction ($\phi=0.5$, purple), $\phi=0.25$ (green) and superior conjunction ($\phi=0$, yellow).}
\label{fig:tau}
\end{figure}

In this section, we focus on the coherence time scales computed in bins centered on phases $\phi=0$, $\phi=0.25$ and $\phi=0.5$ and of width 0.1 each. To study the evolution of the coherence time scale with the parameters, we want to compare the coherence time scale $\tau$ to the amount of time a single clump intercepts the \los to the accreting compact object that is the duration for the X-ray source to cross a clump projected along the \los. We call it the self-crossing time and write it $\tau_{SC}$. Due to the short orbital periods and the high total masses in \hmxbs, the orbital speed can be as high as a few hundreds of kilometers per second. This speed is not negligible compared to the speed of the wind, especially if the acceleration is very gradual (\ie high $\beta$). We compute the local relative speed $\Delta v$ between the orbiting X-ray point source and the densest clumps along the \los:
\begin{align}
\begin{split}
\label{eq:rel_speed}
    \Delta v=& [ v_{orb}^2\left(1-\sin^2 \Phi \sin ^2 i\right)+v_{cl}^2\left(1-\cos ^2 \Phi \sin ^2 i\right) \text{...}\\
    & \text{...} - 2v_{orb}v_{cl}\cos \Phi \sin \Phi \sin ^2 i ] ^{1/2}
\end{split}
\end{align}
where $\Phi=2\pi\phi$, $v_{orb}=2\pi a/P$ and $v_{cl}=v(r=a)$. From this, we deduce the self-crossing time:
\begin{equation}
    \tau_{SC}=2R_{cl}/\Delta v.
\end{equation}
To compute the clump radius $R_{cl}$ which enters this equation, we focus on the clumps which contribute the most to $N_H$, like we did in section\,\ref{sec:dNH_params}: at inferior conjunction and $\phi=0.25$, we compute the clump radius at $r=a$ while at superior conjunction, we compute the clump radius at the point along the \los which is the closest from the donor star. 

Figure\,\ref{fig:tau} provides an overview of the ratios $\tau/\tau_{SC}$ for all simulations. Except for a handful of simulations where clumps are the largest and the scarcest, the coherence time scale matches the self-crossing time within 50\%. This result means that a fairly accurate estimate of the clump self-crossing time and, by then, of the characteristic dimension of the clumps, can be obtained by measuring the full width at half maximum of the autocorrelation function of a column density time serie. Within the ranges of parameters we explored, $\tau_{SC}$ depends on all the 6 dimensionless parameters summarized in Table\,\ref{tab:params} except on the clump mass. It provides a convenient way to distinguish the size and the mass of the clumps, which were intertwined when we analyzed the standard deviation on column density in Section\,\ref{sec:peaktopeak}. 

\section{Observational diagnostics}
\label{sec:diagnostics}

\subsection{Procedure}



The procedure we propose to determine clump properties from the variability of the column density follows the 3 successive steps below.

First, we need to have constrained values for the orbital inclination $i$ and separation $a$, the shape of the velocity profile (\ie the $\beta$-exponent) and the column density scale $N_{H_0}$. This can be done based on the median column density profile and if possible, additional insights (\eg radial velocity profiles or the extent of the stellar Roche lobe). 

Then we can make use of the $N_H$ time series provided by observations, for instance through time-resolved X-ray spectroscopy \citep[\eg][]{Martinez-Nunez2014} or color-color diagrams \citep[][]{Nowak_2011a,Grinberg_2020a}. The clump radius manifests as a coherence time scale imprinted in the signal which can be extracted from the width of the autocorrelation function at half its maximum. It typically ranges from a few $10^{-4}$ to a few $10^{-2}$ times the orbital periods, which means that a time resolution of $\sim$100 seconds is enough to determine the clump radius if a coherence time scale is detected, and to provide constraining upper limits on the size of the clumps otherwise. The relative speed projected on the plane of the sky of the clumps with respect to the X-ray point source is lower near inferior conjunction. Consequently, for systems where the inclination is low enough that we can afford it ($i<\ang{65}$), measures at inferior conjunction should be privileged since they yield longer coherence time scales that are easier to observationally constrain. Otherwise, measures at $\phi\sim0.2-0.3$ are to be preferred. 

Finally, we can use the standard deviation of the column density, $\delta N_H$, to estimate the clump mass based on equation\,\eqref{eq:delta_NH_general}. At inferior conjunction, $\delta N_H/N_{H,0}$ provides a direct measure of the ratio $\sqrt{m_{cl}}/R_{cl}$ whose accuracy is within a factor of 2 in the worst case, when no other parameter ($a$, $i$, $P$ and $\beta$) is known (see Figure\,\ref{fig:sigma}). Since the clump radius $R_{cl}$ entering this ratio is the same as the one deduced from the coherence time scale, the characteristic mass of a single clump can be derived.

\begin{figure}
\centering
\includegraphics[width=0.99\columnwidth]{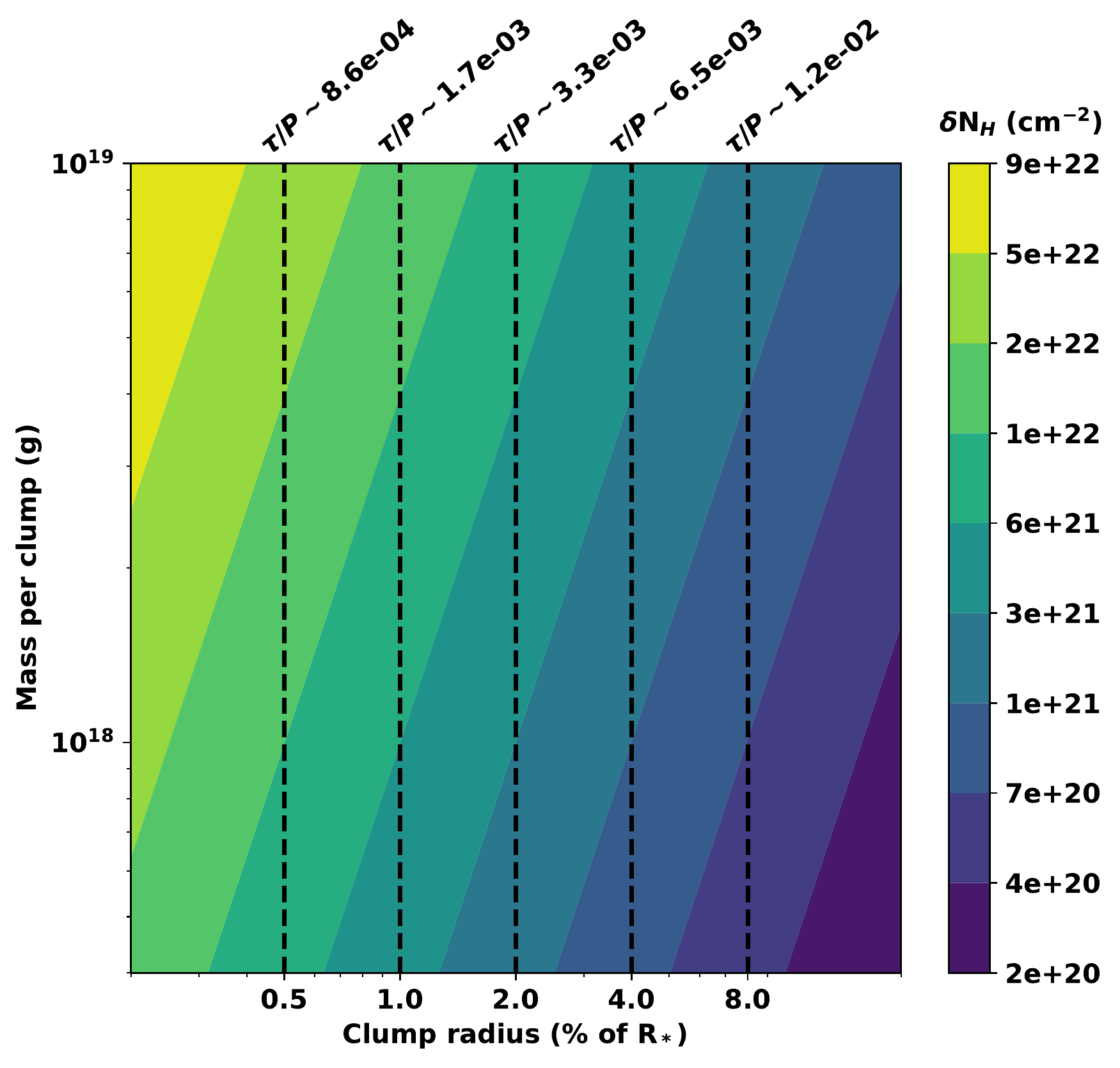}
\caption{Synthetic diagram summarizing the dependence of the coherence time scale $\tau$ and of the standard deviation on the column density $\delta N_H$ on the size of the clumps and their mass, for intermediate parameters ($a=2$, $i=\ang{45}$, $P=24$ and $\beta=2$). We used as normalization units $m_0=10^{24}$g for masses and $N_{H,0}=3\cdot 10^{23}$cm$^{-2}$ for column densities.}
\label{fig:summary}
\end{figure}

The diagram in Figure\,\ref{fig:summary} summarizes the main ideas underlying this procedure for observational diagnostics of the wind micro-structure, showing the relationship between the mass and radius of an individual clump, and the standard deviation and coherence time scale of the column density $\delta N_H$, here calculated for a specific case. For illustrative purposes and to provide masses in physical units, we considered arbitrary scaling units and a simulation with intermediate parameters. The coherence time scales $\tau$ are given in orbital periods and directly map for clump radii. The additional information on $\delta N_H$ enables to constrain the range of possible clump masses using:
\begin{align}
\begin{split}
    m_{cl}\sim & 3\cdot 10^{22}\text{g}\left(\frac{R_{cl,m}}{R_*}\right)^2 \times \text{...}\\
    & \left(\frac{R_*}{20R_{\odot}}\right)^3 \left(\frac{v_{\infty}}{1\,000\text{km}\cdot\text{s}^{-1}}\right) \times \text{...}\\
    & \left(\frac{\dot{M}}{10^{-6}\text{\msun}\cdot\text{yr}^{-1}}\right)^{-1} \left(\frac{\delta N_H}{5\cdot 10^{21}\text{cm}^{-2}}\right)^2
\end{split}
\end{align}
The corresponding porosity length is, for the parameters and mass unit used in Figure\,\ref{fig:summary}:
\begin{equation}
    h\sim 2\% R_* \left(\frac{m_{cl}}{4\cdot 10^{17}\text{g}}\right) \left(\frac{R_{cl,2}}{0.01R_*}\right)^{-2}.
\end{equation}
\red{This small value of the porosity length is typical of an environment with many low mass clumps susceptible to intercept the \los and capture photons.}

\subsection{Comparison to previous results for clumps in HMXBs}
\label{sec:big_clumps}

For clump sizes and masses chosen based on radiative hydrodynamics simulations ($R_{cl}\sim 0.01R_*$ and $m_{cl}\sim$ a few $10^{17}$g, \citealt{Sundqvist2017}), the dispersion of the $N_H$ values around the smooth wind profile is limited. 
The clumps with masses $m_{cl}\sim 10^{22-23}$g and radii $R_{cl}\sim$ a few $R_*/10$ deduced from hard X-ray flares in \sfxts \citep{Walter2007}, which trace the intrinsic emission and thus, the mass accretion rate, are more likely not representative of the capture of a \red{single} clump by the compact object. Instead, the flares could be due either to large scale structures crossed by the accretor \citep[\eg corotating interaction regions,][]{Bozzo2017a} or either to instabilities and gating mechanisms at the outer rim of the \ns magnetosphere such as the propeller effect \citep{Bozzo2008,Parfrey2017}, or within the accretion disk surrounding the black hole. These instabilities could be triggered by a clump, especially in \sfxts \citep{Ferrigno2020}, but involve much larger amounts of mass. For the edge-on \hmxb Vela X-1, \cite{Grinberg2017} carried out a time-resolved analysis of the X-ray spectra and identified two coherent episodes of enhanced absorption of approximately an hour each, which would correspond to clumps of size $R_{cl}\sim 5\% R_*$. However, the amplitude of the $N_H$ increase, of several 10$^{22}$cm$^{-2}$, is too high to be explained by clump-induced stochastic variations. In edge-on systems, these observations could be due to variability in the accretion wake or, when observations are performed at $\phi\sim$0.25 like in \cite{Grinberg2017}, to the formation of transient high density axisymmetric structures around the \ns magnetosphere such as an accretion disk \citep{ElMellah2018,Liao2020}. 


\section{Limitations of the model}
\label{sec:limitations}

\subsection{Model-intrinsic limitations}

\subsubsection{Caveats for edge-on systems}
\label{sec:inclination}

Since the first simulations of wind accretion by \cite{Blondin1990}, the flow in \hmxbs is believed to form, in the orbital plane, a hydrodynamics bow shock which trails the accreting compact object due to the orbital motion. \cite{Manousakis2015c} showed that this dense structure was highly variable due to the X-ray ionizing feedback from the accretor onto the feeding material upstream, which alters the line acceleration as first described by \cite{Hatchett1977} (see Section\,\ref{sec:ionizing_feedback}). We intentionally left large scale density modulations out of the model introduced in the paper so as to isolate the effect of the clumps themselves on the variability of the column density. In order to apply the present model in \hmxbs with high orbital inclination (eclipsing and/or $i>\ang{65}$), observations from superior conjunction ($\phi=0$), or egress if eclipsing, up to $\phi\sim 0.2-0.3$ are relevant to determine the wind micro-structure. In this interval, the accretion wake points away from the observer and does not intercept the \los. Elsewhere, the column density is significantly impacted by the contribution of structures not associated to the clumps. Instead, for intermediate inclinations ($\ang{25}<i<\ang{65}$), data up to inferior conjunction can safely be used, especially if the wind speed is large compared to the orbital speed, leading to a low opening angle of the bow shock around the accretor. If the inclination is low ($i<\ang{25}$), X-ray data at any orbital phase can be used but the weak modulation of photometric and spectroscopic signals with the orbital motion makes quantities such as the orbital period difficult to narrow down. Notice that a low inclination is much less likely than an edge-on configuration since the probability to observe a system with orbital inclination $i$ goes as $\sin i$. 

\subsubsection{X-ray photoionization}
\label{sec:ionizing_feedback}



Wind accretion being less efficient at transferring stellar material to the compact object than Roche lobe overflow, the mass accretion rate in wind-fed \hmxbs is generally moderate, leading to typical X-ray luminosities on the order of 10$^{36}$erg$\cdot$s$^{-1}$. Nevertheless,
the influence of the X-rays emitted from the immediate vicinity of the compact object on the ionization state of the upstream wind can often not be neglected \citep[see \eg][for observational results]{Boroson2003,Miskovicova2016b,Grinberg_2020a}. In addition of altering the wind dynamics \citep[\eg][]{Cechura2015,Sander2017}, the X-rays ionize the material surrounding the compact object and significantly decreases its opacity \citep{Hatchett1977,Ho1987a}. In particular in the soft X-ray range used for absorption measurements in HMXBs, the effects of changing ionization lead to complex, nonlinear changes in observed spectral shape. In general, segments of the \los near the accretor will pass through material that is less absorbent;  the apparent column density inferred from X-ray spectroscopy is then, if a neutral absorber is assumed, smaller than the actual total column density of material and thus also than the column density computed in this paper.

A first correction can be made to the smooth wind profile\,\eqref{eq:NH_smooth} by accounting only for the material outside a region where the ionization parameter\footnote{The ionization parameter is usually defined, after \citet{Tarter_1969a} as $\xi = L_\mathrm{x} / n r^2$ with $L_\mathrm{x}$ the ionizing X-ray luminosity, $n$ the number density of illuminated material and $r$ the distance from the X-ray source. It is often expressed as $\log \xi$, with $\xi$ in CGS units.} $\xi$ is lower than a critical value $\xi_c$. 
However, the exact value of $\xi_c$ will depend on details of the observations used to obtain the absorption measurements from X-ray spectra, for instance the energy ranges considered and underlying spectral degeneracies, and on detailed (micro-)structure and composition of the wind. Only a proper radiative transfer computation  can provide the insights we need on the wind ionization structure to correct  the total column density $N_H$ \citep{Watanabe2006,Mauche_2008a}. However, such simulations are not yet available for complex, dynamic clumpy winds. Interestingly, in \sfxts, the lower X-ray luminosity than in classic \sgxs should limit the impact of photoionization on the column density \citep{Pradhan2017}. Also, the \sgx 2S0114+650 discussed in section\,\ref{sec:analytics_smooth} may, due to its moderate X-ray luminosity, be a good candidate to constrain the total column density. Investigations of the stochastic properties of the column density along the grazing orbit would be a precious element to gain knowledge on the size, mass and onset region of the clumps in the wind of the B1 supergiant donor star in 2S0114+650.


An additional problem may be posed by inhomogeneous and variable X-ray irradiation, such as when considering time scales smaller than the pulsar rotation period in HMXBs with accreting pulsars as compact object. In such systems, changes in X-ray flux can reach a factor of a few within 10--100\,s, resulting in a dynamic illumination of the wind material. Around active galactic nuclei, the resulting effects of the X-rays on the clumps have been studied thanks to radiative transfer computations by \cite{Waters2017}. A similar dedicated investigation in \hmxbs is still required to evaluate how impacted the wind ionization structure can be by the spinning pulsar.


\subsubsection{Eccentric orbits}
\label{sec:ecc_orbits}

For the sake of simplicity, we kept the number of degrees of freedom of our model limited to a minimum and discarded eccentric orbits. However, the present model can straightforwardly be extended to study a specific system where the orbit is suspected to be eccentric, especially when it comes to the fitting of the median $N_H$ profile by the smooth wind solution\,\eqref{eq:NH_smooth}. Although classic \sgxs such as Vela X-1 are largely circularized, \sfxts exhibit more eccentric orbits and would benefit from such an extension. For instance, the $N_H$ profile obtained by \cite{Garcia2018} in the \sfxt IGR J16320-4751 displays a very sharp increase which the authors interpret as due to a grazing orbit. In addition, the flat $N_H$ profile they measure away from this spike can only be reproduced either with a low inclination, incompatible with a grazing orbit, or with an eccentric orbit. They derived an eccentricity of 0.2, possibly indicative of the conditions of formation of the compact companion \citep{Brandt1995}. Even if they are more rare, eccentric orbits can occur in classic SgXBs too: in 2S0114+650 discussed in Section\,\ref{sec:analytics_smooth}, some authors found an eccentricity close to 0.17 \citep{Crampton1985,Grundstrom2007} although \cite{Koenigsberger2006} argued in favor of a zero eccentricity, which could account for the super-orbital period measured in this system.


\subsubsection{Varying clump properties}
\label{sec:diff_cl_properties}

A final limitation to be acknowledged is the unique clump mass and radius at $r=2$ we consider. In reality, numerical simulations indicate various masses and radii for the clumps \citep{Sundqvist2017}. \cite{Ducci2009} considered the impact of a distribution of sizes and masses of the clumps to study the time variability of the mass accretion rate. We avoided introducing new parameters in the present study but such a distribution could be included in our simulations to seed the clumps. Similarly, it would be possible to study in more detail the impact of nonspherical clumps on the column density \citep{Oskinova2012}.

\subsection{Observational challenges and prospects}
\label{sec:obs_chal}

 In the Galaxy, we currently know $\sim$20 wind-accreting \sgx and $\sim$10 SFXTs with \ns compact objects \citep{Martinez-Nunez2017} as well as one wind-accreting \sgx with a \bh, which constitutes a reasonable sample for observational studies. However, next to the model-intrinsic caveats listed above, our theoretical approach and suggested procedure do not yet take into account observational limitations and biases for measurements of $N_H$ on the required time scales. An in detail discussion of these issues is out of scope of this paper, as it would require simulations for different instruments, observational set ups, source properties, and analysis approaches. We thus limit ourselves to giving some first estimates for observational challenges and prospects that could include:

\begin{itemize}
    \item possible cadence of X-ray measurements. In particular. X-ray satellites in low Earth orbit may not be able to provide uninterrupted measurements long enough to obtain meaningful measurements of the coherence time scale. As an example, we assume comparatively small clumps ($R_\mathrm{c} = 0.005\,R_*$ and thus $\tau / P \sim 8.6\times10^{-4}$, cf. Fig.~\ref{fig:summary}) in the black hole SgXB Cygnus X-1 ($P = 5.6$\,d), resulting in a coherence time scale of $\sim$400\,s. This has to be compared to the typical length of an uninterrupted observation with a low Earth satellite such as \textsl{RXTE} of on average $\sim$1.7\,ks and a maximum of $\sim$3.3\,ks, that is only 4 and 8 full coherence cycles, respectively \citep{Grinberg2015}. Using such data for measurements would thus be challenging. On the other hand, instruments on elliptical orbits can potentially continuously observe a source for longer periods, up to $\sim$100\,ks \citep[\eg][]{Haberl1989,Odaka_2013a,Miskovicova2016b,Grinberg2017}.
    
    \item achievable time resolution for reliable absorption measurements. In the above example, we would need a time resolution of $\sim$40\,s to cover the coherence time scale with 10 measurements. The necessary time resolution would increase for larger clumps, toward $\sim$600\,s for $R_\mathrm{c} = 0.08\,R_*$, cf. Fig.~\ref{fig:summary}, and be longer for systems with larger orbital periods. Such measurements are challenging. In the eclipsing HMXB Vela X-1, absorption has been directly modeled at different orbital phases on time scales of $\sim$1\,ks using \textsl{Suzaku} \citep{Odaka_2013a} and \textsl{XMM-Newton}'s EPIC-pn \citep{Martinez-Nunez2014} and on time scales of individual pulsar pulses, that is $\sim$280\,s, using \textsl{NuSTAR} \citep{Fuerst_2014a}. The variability in 4U~1700$-$37 has been studied by \citet{Haberl1989} on time scales of $\sim$1\,ks. Typical formal errors on the measurements listed here will depend on the absolute value of $N_H$, but measurements of $\delta N_H \sim 0.5 \ldots 1\times10^{22}\,\mathrm{cm}^{-2}$ are realistic \citep{Odaka_2013a}. However, direct absorption modeling from spectral fits on shorter time scales seems unrealistic for today's instruments. With current instruments, the variability of absorption may be accessible on even shorter time scales using indirect methods such color-color diagrams \citep[]{Nowak_2011a,Grinberg_2020a}, although exact measurements are yet lacking for this approach. High throughput X-ray missions such as \textsl{XRISM} \citep{XRISM_2020a}, \textsl{eXTP} \citep{Int_Zand_2019a}, and \textsl{Athena} \citep{Nandra_2013a} will enable even better measurements on short time scales and make a high number of fainter sources accessible for short-term absorption variability studies. For \textsl{Athena} high quality spectra of HMXBs can be obtained on time scales as short as 300\,s \citep[\eg][]{Lomaeva_2020a} and it is worth noting that estimates for the minimum time scale for absorption measurements with any of the listed future instruments have not yet been made.
    
    \item systematic uncertainties in the measurements of the absorbing column density introduced by absorption models used in X-ray astronomy and uncertainties in the underlying continuum emission.  This includes effects of cold vs. (lightly) ionized absorbers and different cross-sections and abundances between different models employed (\ie the chosen opacity of the wind), but also known systematic effects such as the degeneracy between the spectral slope of the continuum emission and obtained $N_H$ values for CCD-resolution instruments \citep[\eg][]{Suchy2008}. To our knowledge, there are no systematic studies of such effects for HMXBs yet that go beyond discussion of individual measurements and observations. An estimate of the overall reliability is thus not yet possible and in particular no comparison of such systematic uncertainties to the intrinsic variability of column density as predicted by our model. 
\end{itemize}
Overall, the observational test of the presented model appears challenging, but possible.

\section{Summary}
\label{sec:summary}

Motivated by the pressing need for a better understanding of stellar outflows from hot stars, we attempted to bridge the gap between theory and observations of line-driven winds from massive stars in orbit with a compact object. We designed a model both complex enough to fairly account for results derived by numerical simulations of line-driven winds and simple enough to be of some use to interpret the observational data from \hmxbs. We highlighted how the wind micro-structure could be constrained based on the variability of the column density. We provided a standard procedure to determine the radius $R_{cl}$ and mass $m_{cl}$ of the overdense clumps formed in the wind of the massive donor star. 
In particular:

\begin{itemize}
    \item 
We identified the fundamental parameters of the problem and illustrated how the stellar mass loss rate could be estimated based on the scale of the orbital $N_H$ profile averaged over several orbits.
\item We showed that the standard deviation $\delta N_H$ of the column density could be computed analytically based on the porosity \red{length \citep{Owocki2006}, 
characterized here as the ratio between the clump column mass and the wind mean density.}
The analytical prediction is in agreement within a few percent with the measured values in the simulations based on the model we designed.
\item In systems where the orbit is grazing, the maximum of $\delta N_H$ is expected to occur at an orbital phase when the \los is tangent to the region above the photosphere where the clumps start to form: the larger this orbital phase, the more extended the smooth wind region is. The measurements of the maximum of $\delta N_H$ and the possible deviation of the measured $\delta N_H$ curve from analytical predictions, thus provides a precious observational diagnostics to determine the location of this region. 

\item  the standard deviation, $\delta N_H$,  was found to be a good measure of the ratio $\sqrt{m_{cl}}/R_{cl}$. We also found that the coherence time scale of $N_H$ time series was an excellent tracer of the clump self-crossing time at the point on the \los the closest from the donor star, provided the relative speed between the clumps and the orbiting X-ray point source was accounted for. The coherence time scale is independent of the number of clumps per unit volume, which enables us to estimate the clump radius. Using the standard deviation $\delta N_H$, we can subsequently obtain an estimate on the mass of individual clumps without resorting on arguments using the time variability of the intrinsic X-ray emission.
\end{itemize}
We further discussed the current limitation of our model, both in terms of intrinsic limitations of the current simulation set-up and possible observational challenges. In practical terms, the suggested procedure will likely requires multiple X-ray exposures of a few 10\,ks each of \hmxbs with moderate luminosities and intermediate orbital inclinations, around orbital phases $\phi=0.25$ and at inferior conjunction ($\phi=0.5$) with sensitive X-ray instruments, that allow $\gtrsim$100 individual measurements of absorbing column density in each exposure.

The heuristic model we designed is aimed at systems where the column density is essentially due to unaccreted clumps passing by the \los, but it falls short of explaining several observations in \hmxbs. Noticeably, the amount of mass accreted during X-ray flares is 10$^{19-23}$g, orders of magnitude larger than the individual clump mass obtained by radiative hydrodynamics simulations of the line-deshadowing instability. Also, the clump sizes deduced from the duration of enhanced absorption episodes are a few times larger than expected from first principles computations. This tension between observation and theory deserves to be addressed with additional investigations of the column density in \hmxbs where the intermediate to low orbital inclination guarantees that the variability of $N_H$ is likely due to the clumps and not to large scale structures produced in the orbital plane by tidal forces.

 




With the advent of a new generation of X-ray satellites in the decade to come, time-resolved X-ray high resolution spectroscopy will grant us access to exquisite data. Valuable information on the morphology of the stellar wind in \hmxbs, both at small and large scales, will be within our scope. Other domains will benefit from the upcoming modernization of the observing fleet. In the Active Galactic Nuclei community, several models of parametrized small scale obscurers have been designed to interpret the X-ray observations \citep{Siebenmorgen2015,Honig2017,Buchner2019}. The origin of these structures remains largely unknown though, contrary to their clump counterparts in the wind of massive stars where they naturally form due to the line-deshadowing instability. Combined efforts to understand the propagation of X-rays in these multiphase environments will shed unprecedented light upon the structures surrounding accreting compact objects but also on the accretion process itself, strongly influenced by the ambient medium.

\begin{acknowledgements}
The authors warmly thank the referee for their particularly insightful report that improved both the scientific content and the readability of the papers, in particular for bringing up the possibility to constrain the basis of the clump-forming region from the $\delta N_H$ profiles. IEM wishes to thank Tomer Shenar for information concerning the system 2S0114+650, Antonios Manousakis for fruitful discussions about the underlying computational aspects, Lida Oskinova for insights concerning the modeling part and the members of the X-wind collaboration for the observational consequences of the present analysis. IEM has received funding from the Research Foundation Flanders (FWO) and the European Union's Horizon 2020 research and innovation program under the Marie Sk\l odowska-Curie grant agreement No 665501. VG is supported through the Margarete von Wrangell fellowship by the ESF and the Ministry of Science, Research and the Arts Baden-W\"urttemberg. IEM and JOS are grateful for the hospitality of the International Space Science Institute (ISSI), Bern, Switzerland which sponsored a team meeting initiating a tighter collaboration between massive stars wind and X-ray binaries communities. The simulations were conducted on the Tier-1 VSC (Flemish Supercomputer Center funded by Hercules foundation and Flemish government). \orange{JOS and FAD acknowledge support by the Belgian Research Foundation Flanders (FWO) Odysseus program under grant number G0H9218N.} M.A.L.\ acknowledges support from NASA's Astrophysics Program.
\end{acknowledgements}


\bibliographystyle{aa} 
\begin{tiny}
\bibliography{aa}
\end{tiny}

\end{document}